\documentclass[useAMS,usenatbib]{mn2e}
\usepackage{graphicx}
\usepackage{journals}

\title[Tables of dust properties in exoplanets]
{Tables of phase functions, opacities, albedos, equilibrium 
temperatures, and radiative accelerations of dust grains 
in exoplanets}
\author[J. Budaj et al.]
{J. Budaj$^{1,2}$\thanks{E-mail: budaj@ta3.sk},
M. Kocifaj$^{3}$, R. Salmeron$^{1}$ and I. Hubeny$^{4}$
\\
$^{1}$Research school of Astronomy and Astrophysics,
Australian National University, Canberra, ACT 2611, Australia\\
$^{2}$Astronomical Institute, Slovak Academy of Sciences, 
05960 Tatranska Lomnica, Slovak Republic\\
$^{3}$Faculty of Mathematics, Physics and Informatics, 
Comenius University, Mlynska dolina, 842 48 Bratislava,
Slovak Republic\\
$^{4}$Steward Observatory, University of Arizona, 933 N. Cherry Ave.,
Tucson, AZ 85721, USA}

\begin{document}

\date{Accepted 2015 July 24. Received 2015 July 19; 
in original form 2014 October 3}

\pagerange{\pageref{firstpage}--\pageref{lastpage}} \pubyear{2002}

\maketitle

\label{firstpage}

\begin{abstract}
There has been growing observational evidence for the presence 
of condensates in the atmospheres and/or comet-like tails of extrasolar 
planets. As a result, systematic and homogeneous tables of dust 
properties are useful in order to facilitate further observational and 
theoretical studies. In this paper we present calculations and analysis 
of non-isotropic phase functions, asymmetry parameter 
(mean cosine of the scattering angle), absorption and scattering 
opacities, single scattering albedos, equilibrium temperatures, and 
radiative accelerations of dust grains relevant for extrasolar planets. 
Our assumptions include spherical grain shape,
Deirmendjian particle size distribution,  and Mie theory.
We consider several species: corundum/alumina, perovskite, 
olivines with 0\% and 50\% iron content, 
pyroxenes with 0\%, 20\% and 60\% iron content, pure iron,
carbon at two different temperatures, water ice, liquid water, 
and ammonia.
The presented tables cover the wavelength range of 0.2 to 500 micron
and modal particle radii from 0.01 micron to 100 micron.
Equilibrium temperatures and radiative accelerations assume 
irradiation by a non-black-body source of light with temperatures 
from 7000K to 700K seen at solid angles from 2$\pi$ to $10^{-6}$ sr.
The tables are provided to the community together with a simple 
code which allows for an optional, finite, angular dimension of 
the source of light (star) in the phase function.
\end{abstract}

\begin{keywords}
astronomical data bases: miscellaneous -- opacity -- scattering -- 
planets and satellites: atmospheres -- circumstellar matter.
\end{keywords}

\section{Introduction}

Many classes of astronomical objects have low enough temperature and
high enough density that grains of condensates can be formed. Such 
grains are usually called ``dust'', although some authors use the more 
generic term ``condensates''.  These objects include interstellar or 
interplanetary medium, molecular clouds, the outer regions of active 
galactic nuclei discs, outer atmospheres of cool stars 
(e.g., Mira-type supergiants), supernova remnants, brown dwarfs, 
extrasolar planets, circumstellar, protoplanetary, and debris discs, 
comets, and others.

The optical properties of condensates may not only influence but 
fully govern the emerging spectrum and structure of the object.
Dust can absorb the impinging radiation and convert it directly into 
heating of the grains. This process is called 'absorption'
or 'true absorption' to emphasize that the photon is thermalized.
It is quantified by the absorption opacity. Dust can also scatter 
radiation in a process called 'scattering' without 
being heated. This process is characterized by the scattering opacity.
Furthermore, scattering can be highly asymmetric, a property that is 
described by means of the phase function, which depends on 
the scattering angle (the deflection angle from the original direction
of the impinging radiation).
The most prominent feature is a strong forward scattering for large 
values of $x=2\pi a/\lambda$ where $a$ is particle size (radius) 
and $\lambda$ is wavelength.
Finally, condensation affects the chemical composition of 
the object. It removes the condensed elements from the gas phase 
within the dust cloud and, in the atmosphere, also from the region 
above the clouds due to the rain-out.

This paper focuses mainly on extrasolar planets.
Extrasolar planets (exoplanets) are found at a wide range of orbital 
periods and/or distances from their host stars, ranging from a fraction 
of a day to many years. Their hosts have spectral types from
A-stars to M-stars and down to brown dwarfs. As a result, 
depending on the amount of stellar irradiation, 
the atmospheres of exoplanets may have temperatures from 
a few thousand K to about one hundred K.
At such temperatures condensates are expected to be present in their 
atmospheres or circum-planetary material.
Condensates are usually confined to clouds at a certain depth or 
region according to their condensation temperatures 
\citep[see e.g.][]{ackerman01,allard01,tsuji02,lodders03,sanchez04,
burrows06,lodders06,helling08,morley12,morley14}.
In highly irradiated atmospheres, or at large depths of cooler 
exoplanets, only the most refractory dust species (Ca-Al-Ti bearing 
compounds) can survive. They are overlaid by the optically opaque 
clouds of iron and various silicates.
Higher up in the atmosphere, or at cooler temperatures, finer
sulfide and alkali halide clouds may be present.
Water clouds become dominant at lower temperatures and, 
at even lower temperatures, ammonia clouds are expected.
An entire classification scheme for exoplanets was proposed
by \cite{sudarsky00} based on the expected effects of such clouds 
on their spectra.
 
The above-mentioned dust phase function should not be confused with 
the planet phase function. The later describes the light curve of 
a planet as a function of orbital phase and is modulated by 
the dust phase function.
There are numerous theoretical studies and numerical models of 
the light curves, spectra, planet phase functions, and albedos 
in extrasolar planets including the effect of clouds
\citep{marley99, seager00, sudarsky05, cahoy10, kane11}.
A convenient analytical model for albedos and phase curves
taking into account non-isotropic scattering was presented in 
\cite{madhusudhan12} and \cite{heng14}.
It turned out that hot-Jupiters have very small albedos
in comparison with the Solar System planets and it is quite 
difficult to detect them in reflected light in the optical 
region due to strong absorption in the Na, K 
resonance lines and/or absence of highly reflective condensate
grains \citep{sudarsky00,rowe08,cowan11}.
Nevertheless, recent observations support the presence of clouds 
in their atmospheres.
\cite{pont13} and \cite{lee14} state that transmission spectra 
of HD189733 are consistent with the presence of condensates in 
the atmosphere of this planet. 
\cite{sing13} argues that aerosols, especially corundum, may 
explain the transmission spectra of another hot-Jupiter, WASP-12b.
\cite{wakeford15} examined the effect of various grain 
species and their particle sizes in the atmosphere of a typical 
hot Jupiter (HD189733b) on the transmission spectrum.
They predicted strong enstatite and perovskite features in 
the infrared region.
Also, the atmosphere of super Earth GJ 1214b  must contain clouds
to be consistent with the observed transmission data
\citep{kreidberg14,morley13,howe12}.
\cite{madhusudhan11} concluded that directly-imaged HR 8799 planets 
have thick clouds (thicker than those required to explain data 
for typical L and T dwarfs). On the other hand, \cite{marley12} 
suggested that it is because exoplanets have lower gravities which
moves the clouds and the L to T spectral transition to lower 
effective temperatures than in brown dwarfs.

Forward scattering on dust can be particularly important in 
eclipsing systems during the primary eclipse, when a cool dusty 
object is in front of the main source of light
\citep{budaj11epsaur}.
Recently a few such systems with dusty disks were discovered
\citep{dong14,meng14,rattenbury15}.
In extrasolar planets, this situation is exactly encountered
near the transit.
\cite{dekok12} studied how forward scattering affects
the transmission spectroscopy of exoplanetary atmospheres.
However, dust clouds associated with planets are not necessarily
confined to the atmosphere of the planet.
\cite{rappaport12} discovered a transiting disintegrating exoplanet
KIC12557548b with a comet-like dusty tail extending well beyond 
its Hill radius and featuring a pre-transit brightening.
Recent studies have shown that pre-transit brightening is due to 
strong forward scattering and can be used, together with the color 
dependence of the transit depth, to constrain the particle size of 
dust grains in the comet-like tail, which was found to be 
of the order of 1 micron \citep{budaj13, croll14, bochinski15} 
\citep[see also][]{brogi12,werkhoven14}.
Two other objects of this kind (KOI-2700b, EPIC201637175B) have 
been discovered already by \cite{rappaport14} and \cite{sanchis15}.
The observed tail lengths in these objects are consistent
with corundum or iron-rich silicate dust grains \citep{lieshout14}. 
Finally, scattering on dust in the Earth atmosphere may have important
effects on lunar eclipses or on the Earth transmission spectrum 
\citep{kocifaj94,kocifaj05,palle11,garcia11,garciaetal11},
as well as on the light curve and spectra of Earth 
\citep{tinetti06,stam08}.
A recent summary of the cloud and haze formation in exoplanet 
atmospheres can be found in \cite{marley13}.

Radiative transfer calculations in planetary atmospheres
are very CPU-time demanding. As a result,
in many applications researchers use simplifying assumptions
such as isotropic scattering or analytical Henyey-Greenstein 
dust phase functions given that more exact calculations 
require application of Mie theory or the use of additional 
Mie-scattering codes. In such cases, precomputed dust opacities 
and phase functions may be very convenient and efficient
\citep{hubeny95,burrows06}.
The aim of this paper is to facilitate further research
and provide the community with homogeneous and systematic tables
of dust phase functions, opacities, single scattering albedos,
grain equilibrium temperatures, and radiative accelerations
relevant mainly for the field of extrasolar planets or circum-planetary 
(circum-stellar) material. These tables can be incorporated into more 
complex cloud models, or find additional applications in 
brown dwarfs, protoplanetary discs, comets, and interacting binary 
stars. 

The supplied tables may also be beneficial for analysis
and interpretation of observations from future space missions.
Large area planet transit surveys such as TESS \citep{ricker14}
and PLATO \citep{rauer14} will discover many exoplanets that
will be followed-up by other large telescopes and their atmospheres,
albedos, and phase curves studied \citep{deming09}. 
Superb photometric precision of CHEOPS \citep{broeg13} will 
detect exoplanetary atmospheres and study day-night heat transfer
in hot-Jupiters, and their albedos.
JWST \citep{gardner06,beichman14} will perform transit or occultation 
spectroscopy and direct imaging of planets which will enable chemical 
composition analysis of the planetary atmospheres. It will study 
dust in all sorts of environments including our own solar system 
or proto-planetary disks. Another mission, ECHO, dedicated 
to study the composition of atmospheres of transiting exoplanets 
was proposed \citep{tinetti12,tinetti15}. 

Since there are currently no observational constraints on the shape,
porosity, or aggregate structure of grains in these environments, 
we assume homogeneous spherical particles and Mie theory.
Recently, \cite{cuzzi14} proposed an opacity model which takes into 
account porous aggregates, \cite{mordasini14} developed 
an analytical model for grain opacity in the atmospheres of forming
planets, and \cite{ormel14} presented a method to include 
the evolution of grain size and opacity into the structure
of protoplanetary atmospheres.
The above-mentioned features are not contemplated in the present 
paper and it was not our intention to calculate cloud or atmosphere
models for exoplanets.

This paper is organized as follows.
Section \ref{rip} describes the source of the refractive index.
Section \ref{grid} describes the parameter space covered by 
the calculations.
Sections \ref{calc}, \ref{res} describe the calculations
and properties of the phase functions, opacities, albedos,
equilibrium temperatures, and radiative accelerations of each 
species. 
Section \ref{out} provides a detailed description of the tables,
their format and units.
Sections \ref{conclusions} contains our summary and conclusions.


\section{Complex refractive index}
\label{rip}

The optical properties (cross-sections for scattering, absorption, 
and phase functions) of spherical dust particles composed of a
homogeneous material with a wavelength-dependent refractive index
can be calculated using the conventional Mie theory. 
Here we describe the sources of the refractive index.
The Heidelberg - Jena - St.Petersburg - Database of Optical Constants
is a very convenient starting point \citep{henning99,jager03b}.

{\it Corundum.} 
It is the most common form of crystalline alumina (Al$_{2}$O$_{3}$),
which is $\alpha$ alumina.
It is one of the most refractory condensates that is expected 
to condense from a gas of solar chemical composition.
Corundum is an optically anisotropic (birefringent) material.
We adopted recent IR-mm measurements of its refractive index by 
\cite{zeidler13}.
These are in situ, high temperature measurements, in the range 
of 300-928 K for both ordinary and extraordinary rays. The optical 
properties of corundum were found to be temperature dependent.
Corundum is most important for exoplanets at high temperatures 
since at lower temperatures its effects are superseded by 
more opaque and less refractory species. For this reason,
we adopted the measurements for the highest temperature considered 
in the above mentioned study ($928$ K).
We calculated the optical properties (cross-sections, opacities,
phase functions, asymmetry parameter) independently for both ordinary 
and extraordinary rays. Then we took a weighted sum of them
with weights of 1/3 and 2/3 for extraordinary and ordinary
rays, respectively, as discussed by \cite{zeidler13} to account 
for an averaged crystal orientation.
There are several types of alumina which may differ
in optical properties. Therefore we considered also 
$\gamma$ alumina and its complex refractive index was taken from 
\cite{koike95}. 
Their measurements are at room temperature and cover a broad range
of wavelengths from UV, optical to far IR.
If not otherwise specified, in the next,  we will refer to
$\gamma$ alumina as alumina and $\alpha$ alumina as corundum.

{\it Perovskite.}
Perovskite is calcium titanium oxide (calcium titanate)
with the chemical formula CaTiO$_{3}$.
Similar to corundum, it is one of the most refractory species
and the highest temperature condensates from a gas of solar chemical
composition.
It may form clouds in exoplanets and brown dwarfs \citep{lodders06}
which might be detected in the transmission spectrum 
in the infrared region with instruments like JWST \citep{wakeford15}.
Formation of perovskite is interesting also because it removes
titanium and thus also TiO from the gas phase. 
TiO gas is a major opacity source in the optical region of M dwarfs.
If present, it would be a strong absorber also in exoplanets,
causing temperature inversions - stratospheres 
\citep{hubeny03,knutson08,burrows08} with huge impact on 
the atmospheric structure and spectra.
The complex refractive index of perovskite was taken from
\cite{posch03}.
They made reflectance spectroscopy of a natural perovskite 
crystal and used Lorentz oscillator fits to determine the optimum 
parameters, such as frequencies, strengths, or damping constants. 
Lorentz oscillator parameters were used to compute optical constants 
$n$ and $k$, that in turn predetermine small particle spectra. 
Unfortunately, the dielectric constants obtained this way do not 
cover the whole spectral region and span only wavelengths longer 
than 2 microns.

{\it Pyroxenes.}
The complex refractive index of pyroxenes was taken from 
\cite{dorschner95}, who determined the optical constants from 
reflectance and transmittance measurements. While the reflectance 
data were available for wavelengths ranging from 
$\approx$ 0.2 $\mu m$ to 500 $\mu m$ with Kramers-Kronig analysis 
in 8-80 $\mu m$, the transmittance analysis was possible only 
at wavelengths below 8 $\mu m$ and then above 100 microns.
The quality of data sets was guaranteed by homogeneity tests made for 
the sample during lab experiments.
The optical properties of pyroxene (Mg$_{x}$Fe$_{1-x}$SiO$_{3}$) in 
the optical region are quite sensitive to the amount of iron $1-x$ 
in the mineral. Therefore, we considered three kinds of pyroxenes: 
enstatite (iron free, $x=1$), pyroxene with 20\% of iron ($x=0.8$),
and pyroxene with 60\% of Mg atoms replaced by iron ($x=0.4$).

{\it Olivines.}
Forsterite (Mg$_{2}$SiO$_{4}$) was adopted as an iron-free 
representative of the olivine family of magnesium silicates.
Its optical data were taken from \cite{jager03},
who used sol-gel technique to produce amorphous magnesium
silicates and followed with homogeneity and purity tests based on 
TEM analysis combined with EDX.
We considered also an iron enriched olivine 
(Mg$_{2y}$Fe$_{2-2y}$SiO$_{4}$)
with half-half of Mg-Fe atoms ($y=0.5$) and the data were taken
from \cite{dorschner95}.

{\it Iron.}
In contrast to silicates, iron-like particles show different 
optical characteristics as discussed by \cite{pollack94}, where 
the real and imaginary values of complex refractive index were 
published for wavelengths ranging from 0.1 to 10$^5$ $\mu m$. 
However, for pure iron we have used two data
sources: \cite{johnson74} for $\lambda<0.7$ $\mu m$ and
\cite{ordal88} for $\lambda>0.7$ $\mu m$. In both cases, 
the experimental data were collected at room temperature. 
For the metals, the Drude model -- the classical phenomenological 
approach for the conductivity -- is traditionally used to explain 
the transport properties of electrons.
However, the Drude model is not well suited for the near and 
mid infrared region, thus it may potentially be inaccurate when 
applied to $\lambda>0.7$ $\mu m$.
Nevertheless, \cite{ordal88} concluded that, within 
the experimental error, the non-resonant cavity data can be fit 
with plasma frequency 32,900 cm$^{-1}$ and damping frequency 
213 cm$^{-1}$. Combining their measurements with Kramers-Kronig 
analysis, the frequency-dependent refractive indices were computed 
and provided along with the reflectance data.

{\it Carbon.}
Carbon, mainly in the form of graphite particles, is an important 
ingredient of interstellar dust. It has not been detected
in extrasolar planets, yet.
Nevertheless, we included carbon dust particles into our tables
since carbon grains are potentially abundant refractory 
species and might be present in circumstellar material,
carbon-rich environments, or disintegrating exoplanets.
The complex refractive index of carbon dust may be quite sensitive
to the temperature due to the presence of free charge carriers
at higher temperatures as shown by \cite{jager98}.
These authors measured the optical properties of amorphous carbon 
for a set of temperatures ranging from 400 C to 1000 C for a broad 
wavelength range from 0.2 to about 500 microns. 
As the temperature increased, an increased 
carbonization and graphitization of the sample was observed.
We adopted their measurements at the highest and lowest temperatures.

{\it Water.}
A new update of the water ice optical constants was published 
recently by \cite{warren08}, who revised the imaginary part of 
the complex refractive index in the visible spectrum. The values 
they tabulated are appropriate for the hexagonal crystal form of 
ordinary ice or frozen water at temperatures not far below 
the melting point (the nominal temperature was 266 K).
Regarding liquid water drops, the optical data published by 
the International Association for the Properties of Water and Steam 
(IAPWS) are probably the most accurate, but are available only 
for the real part of the refractive index and valid only for 
a limited spectral range. 
Therefore we have used Segelstein data \citep{segelstein81}, which
cover a much wider range of wavelengths for both real and imaginary 
values of the refractive index. Segelstein data are well accepted by 
the optical community and thus often used in simulating 
the optical properties of liquid water in various environments.

{\it Ammonia.}
For ammonia (NH$_{3}$) at temperature ranging from 77 K to 88 K 
we used \cite{martonchik84}, which is the more comprehensive 
review of the optical constants of NH$_{3}$ ice. However, we 
stress that optical constants for solid ammonia in the cubic phase 
may differ significantly from those in other phases -- as shown by 
\cite{ferraro80}.

\section{Parameter space - grid.}
\label{grid}

\begin{figure}
\centerline{
\includegraphics[angle=0,width=9.5cm]{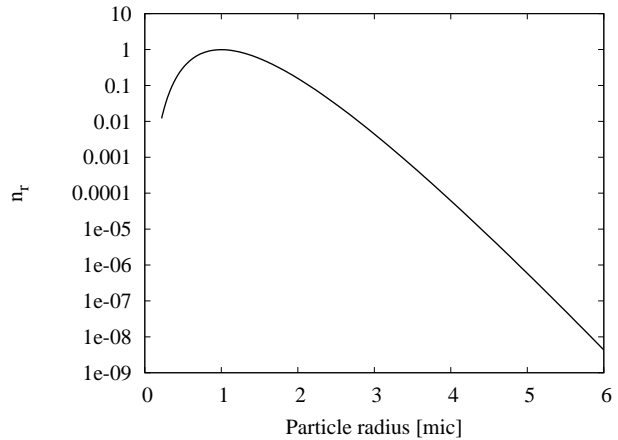}
}
\caption{Deirmendjian particle size distribution
with a modal particle size of 1 micron.}
\label{psd}
\end{figure}

{\it Particle size.}
We assumed homogeneous spherical particles with a poly-dispersed 
Deirmendjian particle size (radius) distribution 
\citep{deirmendjian64}, $n_{r}$, given by:
\begin{equation}
n_{r}=\left( \frac{r}{r_{\rm c}} \right)^{6} e^{-6\frac{r}{r_{\rm c}}}
\end{equation}
where $r_{\rm c}$ is the critical or modal particle radius, where 
the function has its maximum. We truncate the size distribution at 
 $0.22\,r_{\rm c}$ for small particles and at $6\,r_{\rm c}$ 
for large particles. Particular attention is paid to the extended 
tail of larger particles, since their contribution may be
significant e.g. to the Rayleigh scattering or
to the forward scattering peak.
This particle size distribution is shown in Fig.\ref{psd}.
We define the distribution on a very fine integration step 
in order to average the phase functions and to suppress the ripple 
structure that would appear
in the phase function of spherical mono-dispersed particles.
For large particle sizes it is necessary to take into account 
as many as 1600 points in the particle size distribution.
We calculate the dust optical properties for 21 particle size 
distributions with different modal radii ranging from 0.01 
to 100 micron with an equidistant step of 0.2 dex in 
the logarithm of $r_{\rm c}$.
Such particle size distributions can be expected for example
in planetary atmospheres.
Note, however, that although the range of our particle size 
distribution (see Fig.\ref{psd}) spans 
more than 1 dex, the typical full width at half maximum of such 
a distribution is relatively narrow, about 0.5 dex.
Consequently, our results obtained assuming a distribution
with a certain modal radius might be used also as a representative 
values for particles with radii equal to the modal particle radius 
and might be assembled to construct optional, broader, particle size 
distributions if necessary.

{\it Wavelength coverage.} 
Optical properties are calculated for 400 frequency points covering 
the region 0.2 to 500 micron, and spaced equidistantly in logarithm 
of wavelength.
This wavelength range covers most of the spectral energy distribution
(SED) of stars with effective temperatures lower than $\approx 8000$ K.
Not always the entire wavelength region is covered by the measurements
of the refraction index (see the Table \ref{t1} for summary).

\begin{table}
\caption{Summary of the refraction index measurements adopted in
this work.}
\label{t1} 
\begin{tabular}{lllll}
\hline
species    & $\lambda$   &   N  &$\rho^{\rm g}$ & reference   \\
\hline
alumina    & 0.2-400     &  617 &   2.9     &  K          \\
corundum   & 6.7-1e4     & 1500 &   2.9     &  Z          \\
perovskite & 2.0-5843    & 2000 &   4.1     &  P          \\
iron       & 0.19-1.9    &   49 &           &  J\&C       \\
iron       & 0.62-286    &   53 &   7.874   &  O          \\
forsterite & 0.20-949    & 4847 &   3.0     &  J03        \\
olivine50  & 0.20-500    &  109 &   3.0     &  D          \\
enstatite  & 0.20-500    &  109 &   3.0     &  D          \\
pyroxene20 & 0.20-500    &  109 &   3.0     &  D          \\
pyroxene60 & 0.20-500    &  109 &   3.0     &  D          \\
carbon1000 & 0.20-794    &   73 &   1.988   &  J98        \\
carbon0400 & 0.20-518    &   69 &   1.435   &  J98        \\
water ice  & 0.04-2e6    &  486 &   1.0     &  W\&B       \\
water liq  & 0.01-1e7    & 1261 &   1.0     &  S          \\
ammonia    & 0.14-200    &  298 &   0.88    &  M          \\
\hline
\end{tabular}

\medskip
Type of grains, $\lambda$ -wavelength range [$\mu m$], N -number 
of wavelength measurements, $\rho^{\rm g}$ -grain density [$g\,cm^{-3}$], 
reference:
K - \cite{koike95}, Z - \cite{zeidler13}, P - \cite{posch03},
J\&C - \cite{johnson74},
O - \cite{ordal88}, J03 - \cite{jager03}, D - \cite{dorschner95},
J98 - \cite{jager98}, W\&B -  \cite{warren08}, S - \cite{segelstein81}, 
M - \cite{martonchik84}.
\end{table}

{\it Scattering angles.}
Phase functions are calculated for 65 angles ranging from 0 to 180 
degrees. Since the scattering is not isotropic, and may have a very 
strong and narrow forward scattering peak, we adopt a much finer 
step in the forward and backward scattering regimes. 
Angles are symmetric with respect to 90 degrees.

{\it Dust species.}
Optical properties of different condensates are calculated on 
the same grid of 21 particle size distributions and 400 wavelengths.
They are: corundum/alumina, perovskite, pure iron, 
forsterite as an iron-free representative of the olivine family 
of silicates, olivine with 50\% of iron, 
enstatite - an iron-free representative of the pyroxene 
family of silicates, pyroxenes with 20\% and 60\% of iron,
carbon at two different temperatures 1000 C and 400C,
water ice, water liquid, and ammonia.

{\it Star temperatures.}
In addition to inherent dust parameters (dust chemical composition,
particle size) some grain properties (equilibrium grain
temperature, radiative acceleration) depend also on the properties
of the irradiating object.
The most important property of the light source is its SED and 
its angular dimension (i.e. its size and distance).
The SED on the surface of the source is mainly a function of its
effective temperature and, to a lesser extent, also of the surface
gravity, chemical composition, microturbulence and/or other free 
parameters. In case of theoretical models the output SED depends 
also on various simplyfying assumptions and/or input data.
For this reason, we calculated equilibrium grain temperatures
for 13 star/brown dwarf effective temperatures: 
7000, 5800, 5000, 4500, 4000, 3500, 3000, 2500, 2000, 1600, 1200, 
900, and 700 K.
For objects with temperatures $\leq$ 2500 K we assumed
the surface gravity $\log g=5.0$ (cgs), while for hotter objects
we assumed $\log g=4.5$ (cgs). The source is further assumed to have
a solar chemical composition \citep{asplund09,caffau11}. 
We use a single homogeneous grid of synthetic spectra based on 
atmospheric model calculations,
BT-Settl (CIFIST2011-2015, CIFIST2011) \citep{allard03,baraffe15}, 
which covers the entire parameter space. 
These models are 1D, static, with spherical symmetry,
in local thermodynamical equilibrium, with convection and take
into account atomic, molecular and dust opacities.
The spectra list energy flux, $F_{\lambda}$, 
on the surface of the object in $erg\, s^{-1} cm^{-2} cm^{-1}$
from 0.001 to 1000 micron.
This can be easily converted to, $F_{\nu}$, in units of
$erg\, s^{-1} cm^{-2} Hz^{-1}$ via $F_{\nu}=F_{\lambda}\lambda^{2}/c$.
We recall that the effective temperature, $T_{\rm eff}$, of the object
is, by definition, the temperature of the black-body with the same 
total energy output i.e.:
\begin {equation}
\int F_{\nu}d\nu \equiv \sigma T_{\rm eff}^{4}
=\pi \int B_{\nu}(T_{\rm eff})d\nu,
\label{teff}
\end {equation}  
where $\sigma$ is the Stefan-Boltzmann constant and
$B_{\nu}$ is the Planck function.

{\it Solid angles.}
The dust temperature depends also on the solid angle subtended by 
the source of light.
We considered 15 values, roughly logarithmically spaced, 
ranging from $2\pi$ up to $10^{-6}$ sr.
Radiative accelerations are calculated for the same
stellar effective temperatures and particle sizes.
Their dependence on the solid angle becomes insignificant at 
a relatively small distance from the source and thus was not included.

\section{Calculations}
\label{calc}

\subsection{Opacities}
We use a widely available Mie scattering code CALLBHMIE
adapted by B.T Draine. It calls iteratively the Bohren-Huffman 
Mie scattering subroutine BHMIE, \cite{bohren83}, Appendix A. 
It provides $Q_{\rm a}$, $Q_{\rm s}$, and $Q_{\rm e}$, which
are efficiency factors for absorption, scattering, and extinction, 
respectively, for particles with radius $r$.
They are related to the $C_{\rm a}$ and $C_{\rm s}$, the absorption 
and scattering cross-sections, via
\begin {equation}
C_{\rm a}=Q_{\rm a}\pi r^{2},~~~C_{\rm s}=Q_{\rm s}\pi r^{2},~~~
Q_{\rm e}=Q_{\rm a}+Q_{\rm s}.
\end {equation}
It is assumed that the dust grains are composed of 
homogeneous material, have spherical shape, and are placed 
in an environment with 
the complex refractive index of 1-0i.
The code is written in FORTRAN 77. It was modified to calculate 
the average cross-sections for absorption $\overline{C_{\rm a}}$ 
and scattering $\overline{C_{\rm s}}$ for an ensemble of 
particles with an optional size distribution $n_{r}$. 
\begin {equation}
\overline{C_{\rm a}}=\frac{1}{n} \int C_{\rm a}n_{r}dr~~~~~~
\overline{C_{\rm s}}=\frac{1}{n} \int C_{\rm s}n_{r}dr,
\label{c}
\end {equation} 
where 
\begin {equation}
n=\int n_{r}dr.  
\end {equation}  
Subsequently, the absorption  and scattering opacities
$\kappa_{\nu}^{\mathrm{dust}}, \sigma_{\nu}^{\mathrm{dust}}$ of 
condensates in units of $cm^{-1}$ are given by
\begin {equation}
\kappa_{\nu}^{\mathrm{dust}} = \overline{C_{\rm a}}n~,~~~ \quad
\sigma_{\nu}^{\mathrm{dust}} = \overline{C_{\rm s}}n.
\label{ks}
\end {equation}
For practical applications it is more convenient to use the opacities
per gram of dust material, given by
\begin {equation}
\kappa_{\nu,\rho}^{\mathrm{dust}}=\kappa_{\nu}^{\mathrm{dust}}/\rho_{\rm d}
~,~~~ \quad
\sigma_{\nu,\rho}^{\mathrm{dust}}=\sigma_{\nu}^{\mathrm{dust}}/\rho_{\rm d}
\label{kappasigma}
\end {equation}
where $\rho_{\rm d}$ is the density of dust made of particular species
and $\kappa_{\nu,\rho}^{\mathrm{dust}}, \sigma_{\nu,\rho}^{\mathrm{dust}}$
are absorption and scattering opacities, respectively, per gram of 
dust material in units of $cm^{2}\,g^{-1}$. 
The dust density can be expressed as
\begin {equation}
\rho_{\rm d}=\int M_{r}n_{r}dr= \rho^{\rm g} \int V^{\rm g}n_{r}dr=
\rho^{\rm g}\overline{V^{\rm g}}n=\overline{M^{\rm g}}n
\label{rhod}
\end {equation}
where $M_{r}$ is mass of a dust grain of radius $r$.
$\overline{V^{\rm g}}, \overline{M^{\rm g}}$ are the mean volume and
mass of a dust grain, respectively, and
\begin {equation}
\overline{M^{\rm g}}=\frac{\rho^{\rm g}}{n} \int V^{\rm g}n_{r}dr=
\frac{4\pi \rho^{\rm g}}{3n} \int r^{3}n_{r}dr.
\end {equation}
In the above expression, $\rho^{\rm g}$ is the (constant) density of
a grain of dust. 
Substituting equations (\ref{rhod}) and (\ref{ks}) into equation 
(\ref{kappasigma}) one can obtain the following relation between
opacities and cross-sections:
\begin {equation}
\kappa_{\nu,\rho}^{\mathrm{dust}}=
\frac{\overline{C_{\rm a}}}{\overline{M^{\rm g}}}
\label{kappar}
\end {equation}
\begin {equation}
\sigma_{\nu,\rho}^{\mathrm{dust}}=
\frac{\overline{C_{\rm s}}}{\overline{M^{\rm g}}}.
\label{sigmar}
\end {equation}
Using these opacities, the monochromatic optical depth along 
the line of sight $z$ is then given by:
\begin {equation}
\tau_{\nu}= - \int \rho_{\rm d}(z)
[\kappa_{\nu,\rho}^{\mathrm{dust}}(z) + 
\sigma_{\nu,\rho}^{\mathrm{dust}}(z)] dz.
\end {equation}
In the following, the sum of the absorption and scattering 
opacities will be referred to as a total opacity.
Table \ref{t1} also lists the densities of dust grains, $\rho^{\rm g}$,
which we assumed in the opacity calculations.
Note that the density of corundum is $\approx 4 g\,cm^{-3}$ but since 
corundum is most interesting at high temperatures when other 
species are still in the gas phase, we adopted the density of 
corundum close to its melting point which is significantly lower, 
about 2.9 $g\,cm^{-3}$.
On the contrary, the density of ammonia ice rises significantly 
with temperature, but reaches a plateau between 60-100K at about 
0.88 $g\,cm^{-3}$ \citep{satorre13}, which is the value we adopted.
The density of amorphous carbon was also found to increase 
significantly with temperature due to increasing graphitization. 
We adopted the densities from \cite{jager98} for the temperatures
of 400 C and 1000 C which correspond to those adopted for 
the complex refractive index.

\subsection{Phase functions}
We used the same code to calculate the $S_{11}$ element of 
the scattering matrix for uniformly sized particles with radius $r$.
$S_{11}$ describes the angular distribution of the scattered 
light assuming that the incident light is unpolarized.
It is convenient to normalize this angular distribution
such that its integral over the whole solid angle
$d\omega=\sin(\alpha) d\alpha d\phi$ is $4\pi$.
Such distribution is referred to as a phase function,
$p(\alpha,r)$, where $\alpha$ is the scattering angle which 
measures the deflection from the original direction of the photon:
\begin {equation}
\int p(\alpha,r) d\omega=
\int_{0}^{2\pi}\! \int_{0}^{\pi} p(\alpha,r) \sin(\alpha) d\alpha d\phi
= 4\pi.
\label{p}
\end {equation}
The phase function of the population of particles,
$\overline{p}(\alpha)$, with the size distribution $n_{r}$, 
can be calculated in the following way:
\begin {equation}
\overline{p}(\alpha)=\int p(\alpha,r)C_{\rm s}n_{r}dr /
(\overline{C_{\rm s}}n)
\label{p2}
\end {equation}
Since $C_{\rm s}n_{r}$ does not depend on direction, the averaged
phase function $\overline{p}$ has the same normalization. 

The mean cosine of the scattering angle $g$, also known as 
the asymmetry parameter, is calculated from 
the normalized poly-dispersed phase function as:
\begin {equation}
g=\int \overline{p}(\alpha) \cos(\alpha) d\omega ~/
\int \overline{p}(\alpha) d\omega. 
\label{g}
\end {equation}

\begin{figure}
\includegraphics[angle=0,width=8.5cm]{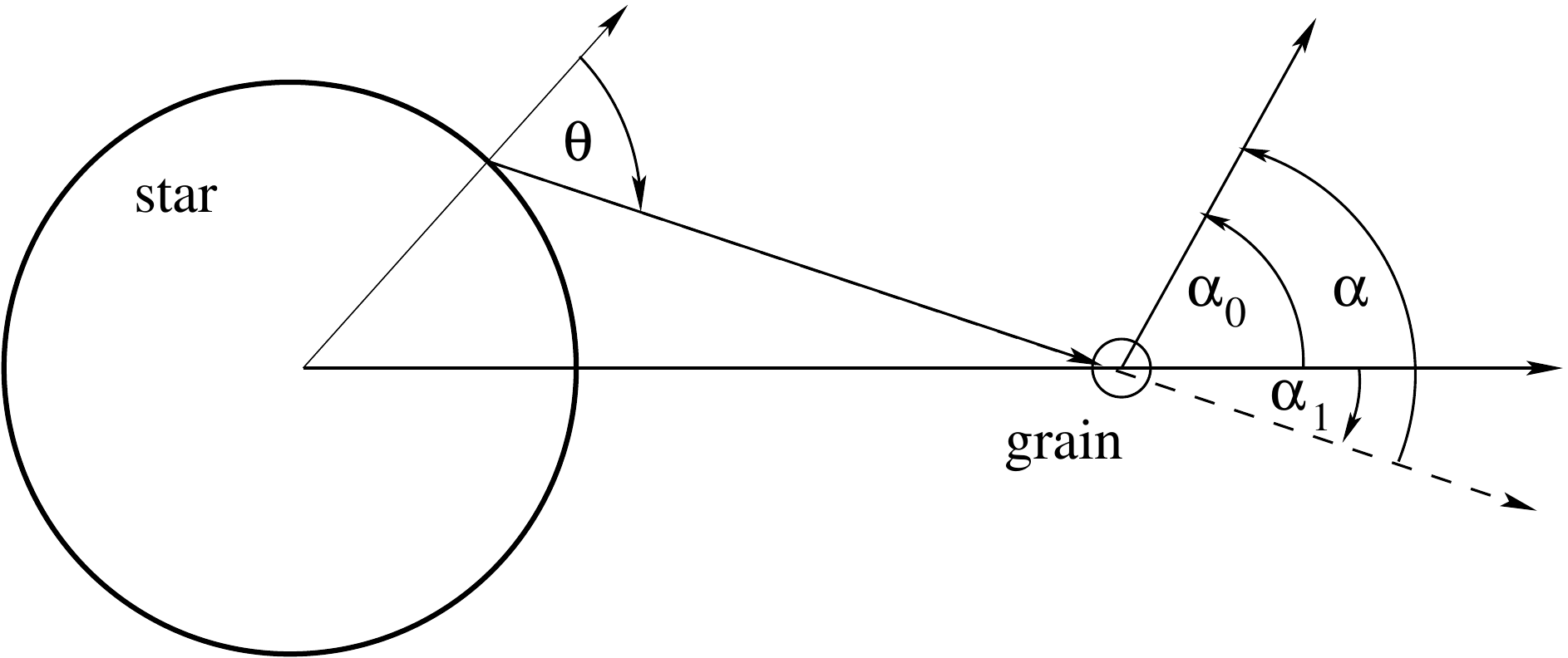}
\caption{
Definition of the geometry and angles.
}
\label{angles} 
\end{figure}

Once the opacities are known, the thermal emissivity of condensates 
(energy per unit time, frequency, volume, and solid angle) 
associated with its true absorption can be calculated as:
\begin {equation}
\epsilon_{\nu}^{\mathrm{th,dust}}=
B_{\nu}(T)\rho_{\rm d}\kappa_{\nu,\rho}^{\mathrm{dust}},
\end {equation}
where $B_{\nu}$ is the Planck function.
The angle-dependent scattering emissivity of condensates into 
a direction $\alpha_{0}$ can be calculated from the scattering
opacity and phase function via the following expression:
\begin {equation}
\epsilon_{\nu}^{\mathrm{sc,dust}}(\alpha_{0})=
\int\!\!\int p[\alpha(\alpha_{0},\alpha_{1}),r] C_{\rm s} n_{r} 
I_{\nu}[\theta(\alpha_{1},\phi)] dr d\omega/(4\pi)
\end {equation}
or 
\begin {equation}
\epsilon_{\nu}^{\mathrm{sc,dust}}(\alpha_{0})=
\int \overline{p}(\alpha)\rho_{\rm d}\sigma_{\nu,\rho}^{\mathrm{dust}}
I_{\nu} d\omega/(4\pi),
\end {equation}
where $I_{\nu}[\theta(\alpha_{1},\phi)]$ is the specific intensity 
of radiation coming from the direction ($\alpha_{1},\phi$) with solid 
angle $d\omega=\sin(\alpha_{1}) d\alpha_{1} d\phi$, 
see Fig. \ref{angles}.
It is usually determined by solving the transfer equation as 
a function of angle $\theta=\theta(\alpha_{1},\phi)$ with respect to
the normal to the surface of the object.
In the following we will write the specific intensity simply as 
$I_{\nu}$.
One can define another averaged phase function, $P_{\rm DA}$
such that:
\begin {equation}
P_{\rm DA}(\alpha_{0})=\frac{\int \overline{p}(\alpha)I_{\nu} d\omega}
{\int I_{\nu} d\omega}=
\frac{\int \overline{p}(\alpha)I_{\nu} d\omega}
{4\pi J_{\nu}},
\label{pda}
\end {equation}
where, $J_{\nu}$ is the mean intensity given by:
\begin {equation}
J_{\nu}=\int I_{\nu}d\omega/(4\pi).
\label{J1}
\end {equation}
This phase function is an average over the source of light.
Since intensity depends on the angle, equation (\ref{pda})
does not guarantee that $P_{\rm DA}$ is normalized, contrary to e.g. 
$p(\alpha,r)$.
However, using $P_{\rm DA}$, the scattering emission can be expressed 
in a very simple way as:
\begin {equation}
\epsilon_{\nu}^{\mathrm{sc,dust}}(\alpha_{0})=
P_{\rm DA}(\alpha_{0})\rho_{\rm d}\sigma_{\nu,\rho}^{\mathrm{dust}}J_{\nu}.
\end {equation}

If we consider a distinct source of light, such as a star for 
example, and a medium surrounding the source and our dust grain, 
which is optically thin, or if we are assuming single scattering events, 
then $P_{\rm DA}$ is an average of the phase function over the surface 
of the star and $\alpha_{0}$ can be conveniently measured
from the ray originating at the centre of the star. 
If the star is far or if it is small, then $P_{\rm DA}$ equals 
$\overline{p}$. However, if the star has a non negligible angular 
dimension on the sky, compared to the characteristic changes in 
the phase function, then one has to take its angular dimension 
into account. The phase function of large grains
or at short wavelengths usually has a very strong forward-scattering 
peak, which can be much sharper than the stellar disc as seen e.g. 
from a typical hot-Jupiter or a close in extrasolar planet.
To take this effect into account one has to split the stellar disc 
into elementary surfaces and integrate the phase function 
over the disc as in the equation (\ref{pda}).
In doing so we assume a quadratic limb darkening of the stellar 
surface:
\begin {equation}
I_{\nu}=I_{\nu}(0)[1-u_{1}(1-\cos\theta)-u_{2}(1-\cos\theta)^{2}],
\end {equation}
where $I_{\nu}(0)$ is intensity perpendicular to the surface of 
the source and $\theta$ is angle between the line of sight and 
a normal to the surface.

\begin{figure}
\centerline{
\includegraphics[angle=0,width=9.5cm]{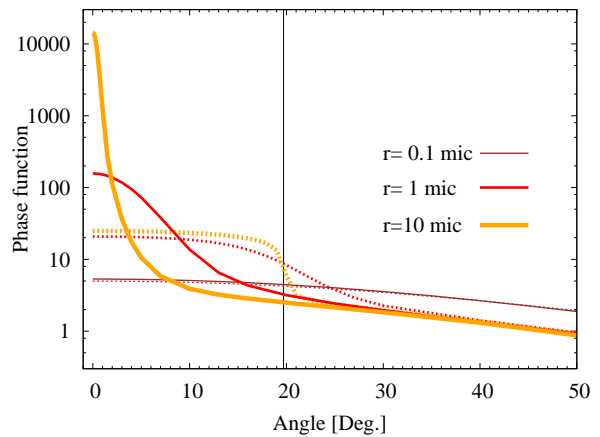}
}
\caption{
Comparison of phase functions assuming a point source of light
(solid lines) versus the phase functions that take into account
the finite dimension of the stellar disc (dotted lines).
Example is for enstatite at 600 nm for different dust particle radii. 
Dotted line for 0.1 micron particles almost coincides with the solid 
line. The vertical line illustrates the angular radius of the stellar
disc of WASP103 as seen from the planet WASP103b
\citep{gillon14,southworth15}.
}
\label{pfbf} 
\end{figure}

One has to be cautious and calculate the phase functions with a very
fine step near the zero angle because of the strong forward 
scattering and, consequently, the disc averaged phase function with 
a very fine step near the edge of the stellar disc.
An illustration of such phase functions, which take into account 
the finite dimension of the source of light, 
is depicted in Fig. \ref{pfbf} for WASP103b. 
For this example we assumed a stellar radius of $R^{*}=1.44 R_{\odot}$
and a planet-to-star separation of $r \approx 4.27 R_{\odot}$ 
\citep{gillon14,southworth15}.
The quadratic limb darkening coefficients $u_{1}=0.36, u_{2}=0.31$ 
were taken from \cite{magic15}.
This results in an angular radius of the stellar disc of WASP103 
of 19.7 degrees i.e. the whole star would appear on 
the planet's sky spanning almost 40 degrees in diameter!
One can see that the finite dimension of the source of light is 
important for larger particles or shorter wavelengths, but has 
almost no effect for smaller particles and/or longer wavelength. 
The Fortran90 code which reads our phase function tables and
calculates such disc averaged phase functions $P_{\rm DA}$ is 
provided together with the tables.
As mentioned above, this code and precalculated $P_{\rm DA}$ phase 
functions are mainly useful in the case that a dust grain 'sees'
a distinct source of light and/or in the optically thin dust regime. 
However, they might be utilized also within the planetary atmospheres,
in which case, one could use $P_{\rm DA}$ for the first scattering 
event and $\overline{p}$ for subsequent scattering events in, 
e.g., Monte Carlo radiative transfer simulations.

\subsection{Albedos}
We calculate and tabulate both the absorption and scattering opacity
as a function of wavelength for each modal particle radius.
These tables and quantities can be used easily to calculate
the single scattering albedos, $\varpi$, for each wavelength and 
modal particle radius via the expression:
\begin {equation}
\varpi_{\nu}=
\frac{\sigma_{\nu}^{\mathrm{dust}}}{\kappa_{\nu}^{\mathrm{dust}}+
\sigma_{\nu}^{\mathrm{dust}}}.
\end {equation}
If not mentioned otherwise, in the text below, by albedo we mean 
the single scattering albedo.

\subsection{Grain equilibrium temperatures}
Electromagnetic interaction with a particle results in heating 
that is ruled by the imaginary part of the complex refractive 
index $m_{\rm i}$ of the particle. 
The larger $m_{\rm i}$, the more light (or electromagnetic radiation) 
is absorbed and consequently transformed to other forms of energy - 
most typically heat.
Let us assume that a particle with an absorption cross-section 
$C_{\rm a}$ is irradiated by a radiation field with an intensity $I_{\nu}$.
The energy absorbed by the particle per unit time is:
\begin{equation}
\frac{dE^{+}}{dt}= \int\!\!\int C_{\rm a} I_{\nu} d\omega d\nu = 
4\pi \int  C_{\rm a} J_{\nu} d\nu,
\end{equation}
where $\omega$ is solid angle. Only angles subtended by the source 
(star) count since $I_{\nu}$ is zero elsewhere.
$J_{\nu}$ is the mean intensity given by equation (\ref{J1}). 
We will assume that this energy is balanced by the thermal radiation
of the particle at a rate:
\begin{equation}
\frac{dE^{-}}{dt}=4\pi \int C_{\rm a} B_{\nu}(T) d\nu,
\end{equation}
where $B_{\nu}(T)$ is the Planck function and $T$ is the temperature 
of the grain. Multiplying both equations by the particle size 
distribution, integrating over the particle size, and dividing by $4\pi$ 
one recovers the radiative equilibrium equation in terms of opacities:
\begin{equation}
\int  \kappa_{\nu}^{\mathrm{dust}} J_{\nu} d\nu 
= \int  \kappa_{\nu}^{\mathrm{dust}} B_{\nu}(T) d\nu.
\label{equil}
\end{equation}
It is common practice to assume a black body radiation for the star 
and neglect its limb darkening, in which case the mean intensity
can be expressed as:
\begin{equation}
J_{\nu}=\frac{\omega}{4\pi}B_{\nu}(T^{*}).
\label{J2}
\end{equation}
In the above expression, $T^{*}$ is the effective temperature of 
the star and $\omega$ is the solid angle subtended by the star:
\begin {equation}
\omega= 2\pi
\left( 1-\sqrt{1-\frac{R_{\star}^{2}}{d^{2}}} \right) ,
\label{omega}
\end {equation}
where $R_{\star}$ is radius of the star and $d$ is the distance 
to the star centre.
However, radiation emerging from stars or brown dwarfs may have 
significant departures from the black-body radiation, especially 
in cooler objects with strong molecular and dust opacities.
Brown dwarf spectra are dominated by the IR water bands which
affect their spectral energy distribution, redistribute the flux
towards shorter wavelengths and interfere with our dust
absorption.
That is why, in the calculations of the left-hand side of 
equation (\ref{equil}), we use the fluxes, $F_{\nu}$, 
from the model atmospheres rather than the Planck function.
In this case, the following equation provides a much better estimate
of the mean intensity (recall equation (\ref{teff})):
\begin{equation}
J_{\nu}=\frac{\omega}{4\pi}\frac{F_{\nu}(T^{*})}{\pi}.
\label{J3}
\end{equation}

Equation (\ref{equil}) can be solved iteratively to obtain the grain 
temperature as a function of stellar temperature and solid angle.
Numerical integration of the equations was performed using 
a simple trapezoidal rule.
Particles of different sizes will have different temperatures
as a result of their different absorption opacities. 
However, since our particle size distribution is relatively narrow, 
we assume that particles within the same distribution have 
the same temperature. Consequently, we calculate and tabulate 
grain temperatures only as a function of the modal radius of particles.
Note that these are equilibrium temperatures an isolated grain 
would have, if there were no evaporation or grain growth. 
In reality, most of the grains 
will not survive at temperatures higher than the condensation 
temperature and will quickly evaporate \citep{kimura02}. 
Also, if gas is present, as e.g. in dense brown dwarf
atmospheres, grains will exchange heat with the gas and both 
constituents settle to the same temperature \citep{woitke03}.
At much lower densities and in the presence of additional heating 
sources, like e.g. in protoplanetary disc atmospheres, condensates 
may have considerably cooler temperature than the ambient gas 
\citep{glassgold04}.

Our grain temperatures cover a broad range of distances (solid angles).
They decline with solid angle. This behaviour is not very
interesting and, therefore to reveal other more interesting features, 
it is convenient to plot the grain temperatures relative to 
the temperature of the grey particle, $T^{\rm grey}$.
The cross-section of grey particle does not depend on the frequency.
In such case equation (\ref{equil}) gives:
\begin {equation}
T^{\rm grey}=T^{*} (\frac{\omega}{4\pi})^{1/4}.
\label{tgrey}
\end {equation}
This quantity will be used below to normalize and visualize 
the grain temperatures.

\subsection{Radiative accelerations}
Our tables can also be used to calculate the radiative acceleration
$a_{\rm R}$ on a particle of certain modal radius and an average mass 
$\overline{M^{\rm g}}$.
The radial component of the acceleration is
\begin {equation}
a_{\rm R}=\frac{1}{\overline{M^{\rm g}}c}\int\!\!\!\int 
\overline{C_{\rm pr}} I_{\nu} d\omega d\nu
\end {equation}
where $c$ is the speed of light and $\overline{C_{\rm pr}}$ is 
an averaged cross-section for radial component of the acceleration 
given by
\begin {equation}
\overline{C_{\rm pr}}=\int[C_{\rm a} A_{1} + C_{\rm s} A_{1}
-\int \frac{p(\alpha)}{4\pi}C_{\rm s} A_{0} d\omega^{'}]
\frac{n_{r}}{n}dr,
\end {equation}
$A_{1}=\cos \alpha_{1}, A_{0}=\cos \alpha_{0}$, 
$\alpha_{1}$ and $\alpha_{0}$ are angles of the incoming and scattered
rays with respect to the radial direction, respectively, $\alpha$
is the phase angle of the scattered beam (see Fig. \ref{angles}).
The integration is performed over all possible directions of 
the scattered photons $d\omega^{'}$, over all particle
sizes $dr$, over all directions of incidence $d\omega$ with intensity 
$I_{\nu}$, and over all frequencies. 
The first term is the contribution from absorption, which
is angle independent since the thermal re-emission of the absorbed
radiation is isotropic.
The second and third therms contain the contribution from 
the scattered photons.
Even for close-in planets, like WASP103b, 
one can assume a point source approximation and that 
$\cos \alpha_{1}\approx 1, \cos \alpha_{0}=\cos \alpha$.
Using equations (\ref{c}),(\ref{p2}),and (\ref{g})
the following expression is obtained
\begin {equation}
\overline{C_{\rm pr}}=\overline{C_{\rm a}} + 
(1 - g) \overline{C_{\rm s}}.
\label{cpr}
\end {equation}
Using the above formula and equations (\ref{J1}) and (\ref{J2})
one gets the following expression for the radiative acceleration
\begin {equation}
a_{\rm R} = \frac{\omega}{\overline{M^{\rm g}}c} \int 
[\overline{C_{\rm a}} + (1 - g) \overline{C_{\rm s}}] 
B_{\nu}(T^{*})d\nu 
\end {equation}
for the black-body approximation, or
\begin {equation}
a_{\rm R} = \frac{\omega}{\overline{M^{\rm g}}\pi c} \int 
[\overline{C_{\rm a}} + (1 - g) \overline{C_{\rm s}}] 
F_{\nu}(T^{*})d\nu 
\end {equation}
when more precise fluxes, $F_{\nu}$, on the surface of 
the source are known.
This can be expressed in terms of opacities using
equations (\ref{kappar}) and (\ref{sigmar}):
\begin {equation}
a_{\rm R} = \frac{\omega}{\pi c} 
\int [\kappa_{\nu,\rho}^{\mathrm{dust}} + 
(1 - g) \sigma_{\nu,\rho}^{\mathrm{dust}}] 
F_{\nu}(T^{*})d\nu.
\label{rad}
\end {equation}

It is very convenient to express the radiative acceleration as 
$\beta=a_{\rm R}/a_{\rm G}$, i.e. relative to the gravitational 
acceleration:
\begin {equation}
a_{\rm G} = \frac{G M_{*}}{d^{2}}.
\label{grav}
\end {equation}
Combining equations (\ref{rad}),(\ref{grav}) and eliminating $d$ using 
$\omega=\pi R_{*}^{2}/d^{2}$ one gets:
\begin {equation}
\beta= \frac{R_{*}^{2}}{G M_{*} c }
\int [\kappa_{\nu,\rho}^{\mathrm{dust}} + 
(1 - g) \sigma_{\nu,\rho}^{\mathrm{dust}}]
F_{\nu}(T^{*})d\nu.
\end {equation}
This $\beta$ is now independent on the distance or $\omega$.
In this way we can calculate and tabulate $\beta$ on the same grid
of stellar effective temperatures $T^{*}$ and particle sizes.
We assumed $M_{*}=M_{\odot}, R_{*}=R_{\odot}$ and the user will need
to adjust our $\beta$ to his object $\beta_{\rm u}$, using 
the following relation:
\begin {equation}
\beta_{\rm u}=\beta \frac{R_{*}^{2}M_{\odot}}{R_{\odot}^{2} M_{*}}.
\end {equation}
The numerical integration of the equations was performed using 
a simple trapezoidal rule, as dicussed next.

\subsection{Numerical methods and precision}
The numerical integration of the equations was performed using 
a simple trapezoidal rule on generally non-equidistant data points.
To assess the robustness of the calculations we carried out a number
of tests. First, we increased the number of integration points in 
the particle size distribution by a factor of two. The opacities 
changed by less than 0.1\%, and the asymmetry parameter by 0.0002.
Second, we extended the small size cut-off radius of the particle size
distribution by a factor of two. The opacities changed by less
than 0.2\%, and the asymmetry parameter by less than 0.0004.
Further, we extended the upper limit of the particle size distribution
by a factor of two to larger particles. This is quite an important 
parameter. It resulted in the maximum change in the opacity of about 
0.3\% and in the asymmetry parameter of about 0.0003. 
The scattering opacity is generally much more sensitive to this upper
cut-off than the absorption opacity since, in the Rayleigh scattering 
regime, it increases with the cube of the particle radius. 
However, the absorption opacity may dominate such 
'relative (fractional)' differences for case of transparent dust 
particles with a very low imaginary part of 
the refractive index and low absolute values of absorption opacities.

We also increased the angular resolution
(decreased the integration step) in the calculations of phase 
functions by a factor of two. The asymmetry parameter changed by less 
than 0.0005 for most of the particle sizes and wavelengths and 
the largest differences in g-values reached about 0.004.  

When extracting fluxes from the model atmosphere tables we
make a simple test and numerically calculate the integral
of the whole SED: $\int F_{\nu}d\nu$. This usually gives a small
departure from the theoretical value of $\sigma T_{\rm eff}^{4}$.
Then, we rescale the fluxes to match precisely 
the theoretical value. Consequently, we use the same integration 
pattern to calculate the dust temperatures and 
accelerations.

Our dust opacities have a short wavelength cut-off at 0.2 micron. 
For the purpose of calculating 
the radiative accelerations and dust temperatures we extrapolate 
them with constant boundary value to 0.001 micron.
If we were to cut them at 0.2 micron this would influence
mainly the hottest stars, that radiate more at shorter wavelengths,
and smaller grains that may have significant opacity increase
at those wavelengths. We found out that the 0.2 micron cut-off
did not change the radiative accelerations of 0.01 and 100 micron 
grains, irradiated by the $T_{\rm eff}=7000$K star,
by more than 11\% and 1\%, respectively, and is negligible for cooler 
models.
Similarly, our dust opacities have a long wavelength cut-off at
about 200-500 micron. We have also checked that neglecting the opacity
at larger wavelengths has a negligible effect on the radiative 
accelerations, since our objects do not have a significant flux at 
those wavelengths.

However, this long wavelength opacity cut-off may not be negligible
in the calculations of the dust temperatures. 
If the dust temperatures are low, the integral on the right hand 
side of equation (\ref{equil}) may be underestimated causing 
artificially high dust temperatures.
For a 500 micron cut-off all grain temperatures that are less than 
about 30 K will be uncertain by more than 2-3\%.
For ammonia and iron, which have about 200 micron opacity cut-off, 
such uncertainty is already noticeable at dust temperatures which 
are less than about 60K.

\section{Output - description of tables}
\label{out}
We present our calculations in the form of four separate files for 
each dust species.

The first file, named e.g. iron\_opac\_all, deals with opacities.
It contains blocks (separated by a blank line) and columns. 
Each block refers to a particular modal particle radius $r_{\rm c}$. 
It has five columns:
(1) $\log(r_{\rm c})$ in micron, (2) wavelength in micron, 
(3) scattering opacity and (4) absorption opacity in $cm^2/g$ 
(per gram means per gram of iron condensates), and 
(5) the mean cosine of the phase function. 

The second file, named e.g. iron\_phase\_all, deals with phase 
functions. It contains several blocks,
each block starts with $\log(r_{\rm c})$ in micron and refers to 
a particular modal radius. Each block has several sub-blocks, 
which apply to a given frequency and are preceded with its 
value in Hz, the mean cosine value, scattering and absorption
opacities in $cm^2/g$ (per gram means per gram of particular 
condensates). Each sub-block has two columns: phase angle in degrees
and phase function normalized to $4 \pi$.

The third file, named e.g. iron\_temp, deals with equilibrium 
temperatures. It also contains several blocks. 
Each block refers to a particular temperature of the star and modal 
particle radius. There are four columns:
stellar effective temperature [K], $\log(r_{\rm c})$ in micron,
solid angle subtended by the star in steradians, and
grain equilibrium temperature [K].

The fourth file, named e.g. iron\_radacc, contains radiative 
to gravitational acceleration ratio, $\beta$, (3-rd column, 
dimensionless) as a function of the particle size 
($\log(r_{\rm c})$ in micron, 2-nd column) and the stellar effective 
temperature (in K, 1-st column).
It assumes that the mass and radius of the irradiating object
are $M_{\odot}, R_{\odot}$, respectively.

Opacities were to zero and the phase functions to one in 
the wavelength regions where the refraction index measurements 
were not available.
The provided 'readme' file contains the description of the tables 
and references to the original refractive index measurements
of individual species.
There is also a computer code DISKAVER.F90 in Fortran90 which 
calculates the disc averaged phase functions.
 
The behaviour of the phase function, opacity, mean cosine, 
albedo, grain equilibrium temperature, and $\beta$ 
is captured in Figs. \ref{alumina} to \ref{ammonia}.

The dust phase function is plotted using logarithmic scale
for 1 micron particles.
These phase functions indeed exhibit strong departures from 
isotropic scattering with very pronounced forward scattering peak.
For large particles, this feature is proportional to 
the cross-section and rapidly increases with the square of 
the particle size. Towards low $x=2\pi a/\lambda$ values 
(small particles and/or long wavelengths) forward scattering 
decreases and backward scattering increases, the phase functions 
approach the dipole phase function, $3(1+\cos \alpha)/4$, and thus
have a similar behaviour to Thomson or Rayleigh scattering 
(with some exceptions such as iron, see below).
The asymmetry parameter reflects the behaviour of the phase functions.

Absorption and scattering opacities as well as
the total opacity (absorption plus scattering) and albedo are 
also plotted using the logarithmic scale.
Particles relatively small compared to the wavelength, with
size parameter $x<<1$, (see e.g. Chap.5 of \cite{bohren83})
show Rayleigh scattering pattern with the scattering efficiency 
proportional to $\sim x^{4}$. 
This results in a $\lambda^{-4}$ wavelength 
dependence and a $r^{4}$ particle-size dependence.
Consequently, the scattering cross-section is proportional
to $r^{6}$ and the opacities per gram follow a $r^{3}$
dependence on particle size. 
If particles are large compared to the wavelength, i.e. $x>>1$, 
$Q_{\rm e}$ approaches a value of about 2 and both absorption
and scattering opacities become noticeable grey.

Albedos are generally high in the optical region.
However, notice that small particles at longer wavelength are in 
the Rayleigh scattering regime, which is not very effective.
As a result, the albedo in this parameter space drops by many orders 
of magnitude to about $10^{-10}$. In the figures, we deliberately 
saturate this regime and assign a dark brown color to albedos 
smaller than $10^{-4}$.
The grain temperatures decrease with decreasing solid angle.
For this reason they are plotted relatively to the grey temperatures, 
$T/T^{\rm grey}$, (see equation (\ref{tgrey})) for a solar type star 
with $T^{*}=5800$K on a grid of modal particle sizes and solid angles.
As the opacities of most large particles become progressively more grey
this results in more grey albedos and equilibrium temperatures.

$\beta$ values are plotted as a function of particle size
for 13 star/brown dwarf temperatures (see Sec.\ref{grid}).
Generally, radiative accelerations can exceed gravity for
hotter stars but the ratio decreases with decreasing object 
temperatures.
The $\beta$ values all show a pronounced peak at 
a certain particle size.
Assuming a black-body radiation of the object, this peak
would be at lower grain radii for hotter stars and would move 
towards larger particles for cooler objects. This is due to 
predominant scattering and its $\lambda^{-4}$ dependence for 
small particles together with strong fluxes at the shorter 
wavelengths for hot objects. 
In the cooler objects, the fluxes move to longer wavelengths and 
so does the Rayleigh regime for larger particles, and this enables 
larger particles to pick-up more momentum. 
We observe that this trend also holds for stars without
the black-body assumtion, i.e. assuming 
synthetic spectra from atmospheric models. However, it stalls 
and may even get reversed in objects cooler than about 1600K.
This is associated with the L-T transition. In T-type brown dwarfs 
the opaque silicate dust clouds settle into deeper layers, 
atmosphere cleares, strong molecular absorption in the infrared 
region pushes the flux via shorter wavelengths reversing the J-K 
color index.
Generally, such 'non-black-body' radiative accelerations of larger 
particles, that tend to have more grey opacities, do not differ 
much from the 'black-body' accelerations, with the exception of water 
and ammonia. However, such 'non-black-body' accelerations 
of 0.1-0.3 micron particles around cool L,T type objects may often be
by a factor of 3-5 higher than their 'black-body' counterparts.
In the next section we briefly summarize the most striking features
for different species, which may be used to identify that 
particular species in the spectra/photometry of exoplanets.

\section{Results}
\label{res}

{\it Alumina/Corundum:}
(Figs.\ref{alumina} and \ref{corundum}) 
absorption opacities of smaller particles show a strong 12 micron 
peak \citep{koike95,zeidler13} which turns into a depression
for larger particles \citep{koike95,zeidler13}.
Scattering shows a dip at 8-9 $\mu m$, a peak at 12 $\mu m$
for smaller particles, and a relatively flat behaviour for larger
particles. As a consequence we see a strong 12 micron feature
in the total opacity of small particles but an enhanced albedo
in the 10-30 micron region for larger particles.
This is accompanied by a sharp drop in the mean cosine 
of the scattering angle $g$ at about 10 $\mu m$ for larger particles.
Overall $g$ is positive and approaches unity for larger particles 
and $\lambda<10$ $\mu m$.
One can see strong forward scattering but weak backward scattering 
features.
Alumina has moderate absorption in the optical regime
and its equilibrium temperatures approach that of the grey absorber.
Smaller particles, when irradiated by a solar type star with solid 
angle $10^{-1}-10^{-3}$, tend to have about 50\% higher temperatures
than the grey absorber.
Large crystallized corundum particles are quit reflective and have 
higher albedo the IR -far IR regime than $\gamma$ alumina.

{\it Perovskite:}
(Fig.\ref{perovskite}) has a strong absorption feature
at 14.1 micron and weaker features at 21.0, 23.8, 33, and 62.3 micron 
\citep{posch03}.
Its albedo is very high in the near-IR region and remains so 
for large particles also at the longer wavelengths.
Significant drops in the albedo are observed near the above mentioned 
absorption features.
Mean cosine of the phase function reaches negative values at 
longer wavelengths and for larger particles which indicates 
significant back-scattering properties. 
On the other hand, highly positive peak g-values are found at 
11 and 20 micron for large particles. 

{\it Forsterite:}
(Fig.\ref{forsterite}) absorption opacities of forsterite show 
the typical double peak at 10 and 18 micron, which is due to 
stretching and bending of Si-O molecular bond in silicates. 
Scattering opacities generally dominate in the visible region
and become progressively more important at longer
wavelengths as the particle size increases. Smaller grain sizes
exhibit the $\lambda^{-4}$ dependence typical of the Rayleigh
regime. These opacity properties are well known and we will now
focus on the behaviour of the phase functions.
The mean cosine of the scattering angle is positive,
which indicates predominance of the forward scattering.
Forward and backward scattering increase towards shorter 
wavelengths or larger particles.
As the particle size increases backward scattering first disappears
at shorter wavelength and then reemerges and expands to progressively 
longer wavelength.
Phase functions also show strong wavelength dependence in the vicinity
of 10-18 $\mu m$ features.
Forsterite absorbs little flux in the optical region which results
in its equilibrium temperatures for solar type stars being 
significantly lower than those of the grey absorber. 
Its albedo is very close to unity in the optical and NIR regions
but large particles remain reflective also in the far-IR and 
sub-mm region. Broad absorption bands at 10-20 micron significantly 
reduce its albedo.

{\it Olivine:}
(Fig.\ref{olivine}) with 50\% of iron has considerably stronger
absorption in UV, visible, and NIR region than forsterite for smaller
and medium size particles.
Small olivine particles have 
slightly higher total opacity in the NIR region. 
The 10 micron feature is sharper, especially its blue edge, while 
the 20 micron feature is slightly broader.
The albedo of olivine is significantly smaller than that of forsterite
in the whole optical and NIR region. Consequently, the equilibrium 
grain temperatures of medium to small particles are several times 
higher than those of forsterite. Larger olivine particles have 
temperatures similar to the grey particles.
This is a consequence of the imaginary part of the refractive index of
olivine, which is much higher in the optical and NIR regions.
Radiative accelerations of the small olivine particles are 
significantly higher than those of forsterite and the maximum is less
pronounced. The forward scattering of 1 micron particles in 
the optical region is narrower while back scattering peak diminished 
and, consequently, the asymmetry parameter of medium size grains is 
larger.

{\it Enstatite:}
(Fig.\ref{enstatite})
behaves in a similar way as forsterite.
Its absorption opacity in the visible and near infrared regions is 
small, even smaller than that of forsterite.
Consequently, its equilibrium temperatures are considerably below
the grey absorber and even lower than those of forsterite.
The phase functions and mean cosine of the phase angle show
even more pronounced wavelength dependence in the 4-20 $\mu m$ region.
The albedo of enstatite has similar behaviour to that of forsterite.
Both iron-free silicates, forsterite and enstatite, show a pronounced 
peak in $\beta$ values with relatively low radiative accelerations, 
especially for small grains, when compared to other species.

{\it Pyroxenes:}
(Figs.\ref{pyrmg80} and \ref{pyrmg40})
even a small amount of iron considerably raises the imaginary part 
of the refractive index in the UV, visible, and near IR regions
\citep{dorschner95}. This raises the absorption opacity
and equilibrium grain temperatures.
Particularly small grains, at relatively close distance 
to a solar type star, with a 20\% amount of iron, 
may become almost 2 times hotter relative to the grey absorber.
Larger particles can still remain relatively cool at larger distances. 
Its albedo shows similarities to iron-free silicates but we start 
to see an enhanced absorption in the UV for larger particles and 
accompanied reduction of the albedo. 
This trend gets even more pronounced and spreads to longer wavelengths 
and smaller particles if the amount
of iron is increased to 60\%.
The presence of iron has also an effect on the phase functions and 
increases the mean cosine of the scattering angle for 
$\lambda<4$ $\mu m$. Backward scattering is apparent in visible and 
near IR regions, however, it diminishes for large particles in UV 
region and this trend is more pronounced with higher iron content.

{\it Iron:}
(Fig. \ref{iron}) contrary to silicates, iron has considerable 
absorption opacity in the visible and IR, which generally smoothly 
declines towards longer wavelengths and larger particles. 
Its absorption and scattering opacities are almost featureless (grey) 
apart from the Rayleigh scattering regime for smaller particles.
The mean cosine of the scattering angle increases with the particle 
size and towards shorter wavelenths. It may have significant 
negative values, especially at longer wavelengths, 
which is due to a very pronounced backward scattering peak.
For larger particles, forward scattering overcomes the backward
scattering at shorter and progressively at longer wavelengths.
These features can be used to identify iron in exoplanets.
Highly effective absorption properties in the optical and IR regions
heat iron grains to significantly higher temperatures as expected for 
the grey absorber. Especially, small iron grains at large distances 
from solar-type stars can get heated to temperatures more than 
a factor of 3 higher than the grey absorber.
As a result of its opacity behaviour, large iron particles
are efficient at scattering in the IR and far IR regime.
However, compared to silicates, iron particles (especially small 
ones) have quite low albedo in the UV and -visible region.
$\beta$ values peak at slightly smaller grains than most of 
other species. Small iron grains will acquire significantly 
higher radiative accelerations than other species when irradiated 
by stars.
This is due to high iron opacities in the optical as well as strong
back-scattering properties which push the particles away from 
the source. On the contrary, larger iron grains tend to have 
slightly smaller $\beta$ values for all source temperatures.

{\it Carbon:}
(Figs. \ref{carbon1000} and \ref{carbon0400})
differs from most of the species but some features resemble those of
pure iron. 
Carbon at higher temperatures and higher graphitization, 
has a total opacity which smoothly declines 
towards longer wavelengths and larger particles. Apart from the slope, 
the dominant 'feature' is the fact that it is almost featureless.
There is a drop in the absorption and scattering opacities
of small particles in the UV.
The most striking characteristic is that its albedo in the optical
and IR regions is much smaller than that of other species, even 
smaller than that of iron. However, the albedo of large particles at 
far-IR and sub-mm wavelengths is approaching unity.
The asymmetry parameter is mostly positive and smoothly increases 
towards the shorter wavelength. However, it becomes negative for 
larger particles in the sub-mm region, a feature indicating stronger 
backward scattering but not as pronounced as in iron. 
The phase functions lack any abrupt features too.

Carbon at cooler temperatures has a steeper gradient of the total 
opacity in the IR region but more shallow in the far-IR and sub-mm 
region \citep{jager98}.
Its albedo features an enhanced band traveling from medium size 
particles at NIR region towards larger particles at sub-mm region
caused by an enhanced scattering at the blue edge of the Rayleigh 
scattering regime.
Apart from this feature, the opacities are almost perfectly grey 
and featureless. Contrary to its behaviour at hotter temperatures,
the asymmetry parameter is everywhere positive and increasing smoothly
towards larger particles and shorter wavelengths. 
Dust temperatures of larger particles and closer to the solar type
star are slightly cooler while smaller particles which are farther 
from the star are significantly hotter than the grey absorber.
Small carbon grains acquire the highest radiative accelerations of 
all the species studied here. 

{\it Water ice:}
(Fig. \ref{waterice})
absorbs very little in the visible but has strong absorption and scattering
features in the IR, particularly at about 3 micron, and shallower features
at about 12 and 40-50 $\mu m$. This is accompanied by the variability
of phase functions at these wavelengths.
Mean cosine values of larger particles show extra particle size sensitivity 
in the 3-10 $\mu m$ region and is larger than zero.
However, broad back scattering feature is present in the visible region.
Its extremely low absorption in the optical 
causes the grains to be significantly cooler than those of the grey
absorber for hotter stars. Its high absorption in the IR, however,
might heat smaller grains at large distances to become 
hotter than grey absorber when irradiated by cooler brown dwarfs.
The albedo of water ice is very high, essentially 1, for all particles
from UV to NIR. Large ice particles are slightly more reflective than 
liquid drops in the far IR and sub-mm regimes. In the IR one can see 
the water absorption bands which lower its albedo.

{\it Water liquid:}
(Fig. \ref{waterliq})
has similar features as ice. It has slightly more pronounced
absorption features, e.g. a nib at about 6 $\mu m$, and slightly 
higher UV and blue absorption.
Mean cosine values of larger particles show enhanced particle size 
sensitivity in the 3-10 $\mu m$ region.

{\it Ammonia:}
(Fig. \ref{ammonia})
has very sharp and strong absorption and scattering features
at about 3, 9.4, 26, 70, 110-140 $\mu m$, which may interfere with 
the watter and silicate features. 
Its absorption in the visible is much stronger than that of watter.
Apart from forward scattering, its phase functions exhibit
also back scattering at the shorter wavelengths and strong
variability with wavelength at the above mentioned spectral regions
which can help to identify ammonia in reflected light.
The equilibrium grain temperatures of ammonia are similar to
those of water but larger particles tend to follow the trend of 
the grey absorber while smaller particles at medium distances are even
cooler.  
Overall, the albedo of ammonia has similar characteristics as that of 
water ice. A number of absorption features can be seen which
reduce its albedo in the IR and far IR regions.
Medium to large ammonia grains also experience strong radiative push
compared to other particles.

\section{Conclusions}
\label{conclusions}
We have calculated phase functions, asymmetry parameters, 
absorption and scattering opacities, single scattering albedos, 
equilibrium dust grain temperatures, and radiative accelerations 
for about a dozen of the species most commonly encountered in 
the extrasolar planet environment. Mie theory and spherical
homogeneous grains are assumed.
The optical properties are integrated across a poly-dispersed 
Deirmendjian particle size distribution with different modal 
particle radii. These distributions are relatively narrow.
Consequently, they might be understood as single sized particles
with the radius equal to the modal radius or used to assemble and 
study an optional broader particle size distribution, if necessary.
These dust properties are presented in the form of tables as 
a function of wavelength and modal particle size. 
Equilibrium grain temperatures are a function of modal particle size, 
host star effective temperature, and solid angle subtended by the star.
Radiative accelerations are tabulated as a function of modal particle
radius and host star temperatures.
Dust opacities are deliberately calculated per gram of the dust 
material so that the user has more flexibility and can place these 
dust particles into any environment independently on the gas to dust 
ratio.
Tables will be maintained, potentially expanded, and available for 
download from:
\begin{verbatim}
http://www.ta3.sk/~budaj/dust
\end{verbatim}
There is also a code which allows the user to take into account,
an optional, finite dimension of the source of light (star)
and calculate grid of disc-averaged phase functions.

\section*{Acknowledgments}
The authors would like to thank Simon Zeidler and Harald Mutschke
for valuable consultations about the optical data as well as
Hannah Wakeford for comments about the manuscript.
JB and RS acknowledge funding by the Australian Research Council 
Discovery Project Grant DP120101792.
MK acknowledges support by the Slovak National Grant Agency 
VEGA grant No. 2/0002/12 and JB also partial support by VEGA 2/0143/14.

\bibliography{budaj}{}
\bibliographystyle{mn2e}

\appendix
\section{}

\begin{figure*}
\vspace{-4mm}
\centerline{
\includegraphics[angle=0,width=8.cm]{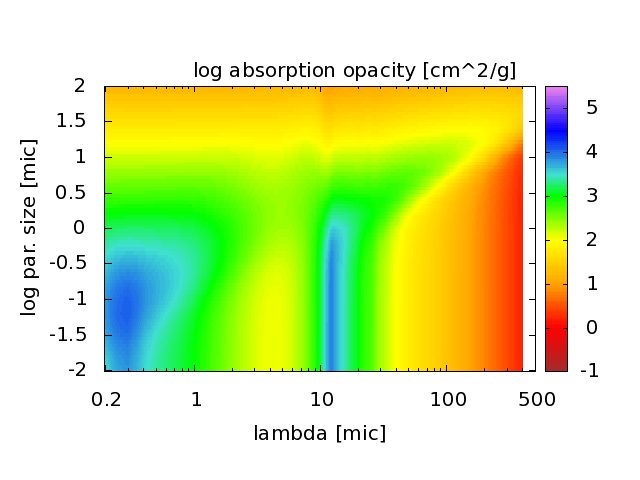}
\includegraphics[angle=0,width=8.cm]{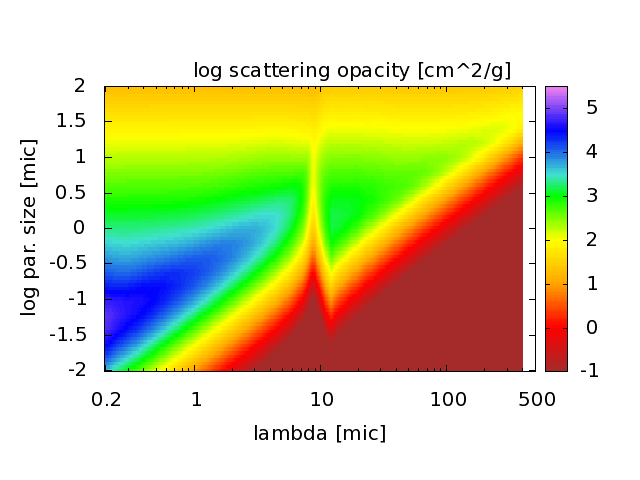}
}
\vspace{-5mm}
\centerline{
\includegraphics[angle=0,width=8.cm]{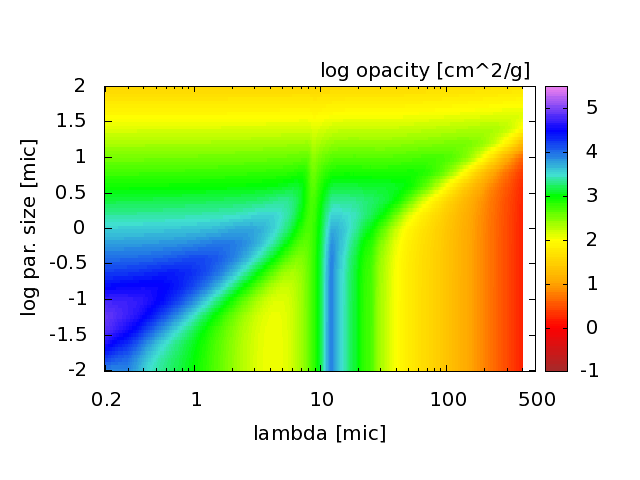}
\includegraphics[angle=0,width=8.cm]{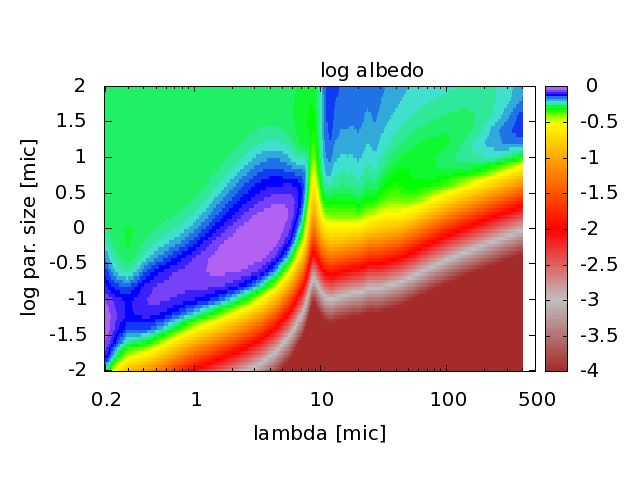}
}
\vspace{-5mm}
\centerline{
\includegraphics[angle=0,width=8.cm]{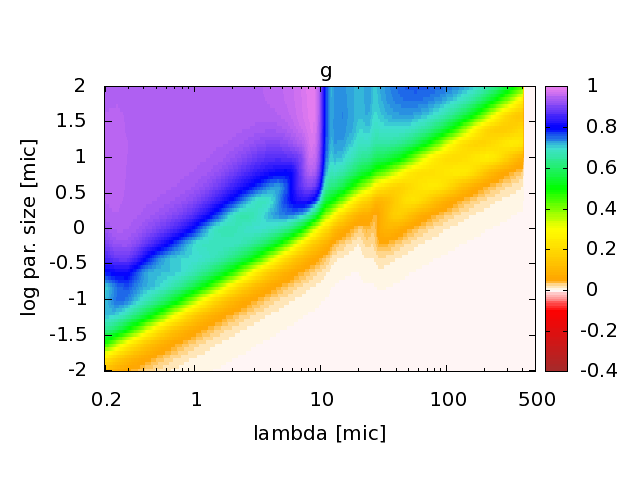}
\includegraphics[angle=0,width=8.cm]{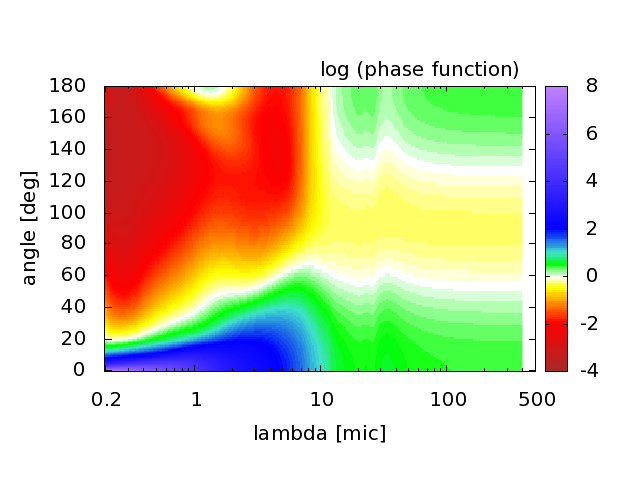}
}
\vspace{-5mm}
\centerline{
\includegraphics[angle=0,width=8.cm]{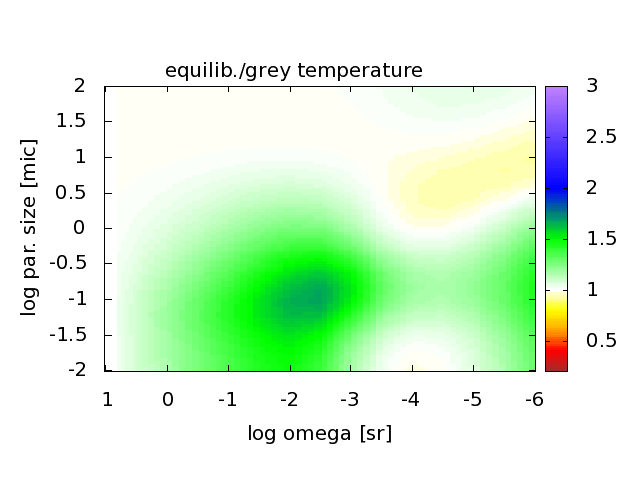}
\includegraphics[angle=0,width=7.5cm]{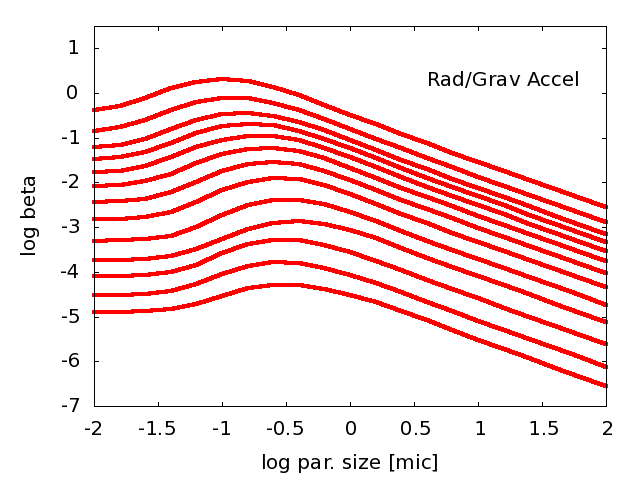}
}
\vspace{-2mm}
\caption{Alumina. 
Very-top: absorption and scattering opacities.
Top left: total opacity (scattering plus absorption).
Top right: single scattering albedo.
Middle left: mean cosine of the scattering angle.
Middle right: phase function for modal particle radius of 1 micron.
Bottom left: Grain temperature relative to the grey absorber
for a solar type star with $T^{*}=5800 K$.
Bottom right: Radiative acceleration relative to the gravity of 
the central object for 13 stellar temperatures, from the top: 
7000, 5800, 5000, 4500, 4000, 3500, 3000, 2500, 2000, 1600, 1200, 
900 and 700 K.
}
\label{alumina} 
\end{figure*}

\begin{figure*}
\centerline{
\includegraphics[angle=0,width=8.cm]{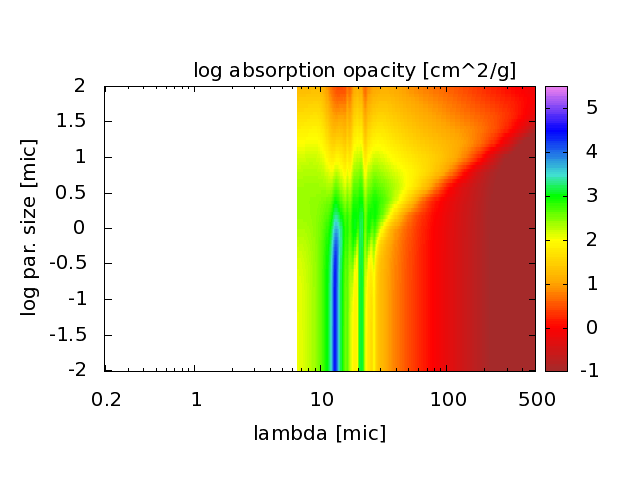}
\includegraphics[angle=0,width=8.cm]{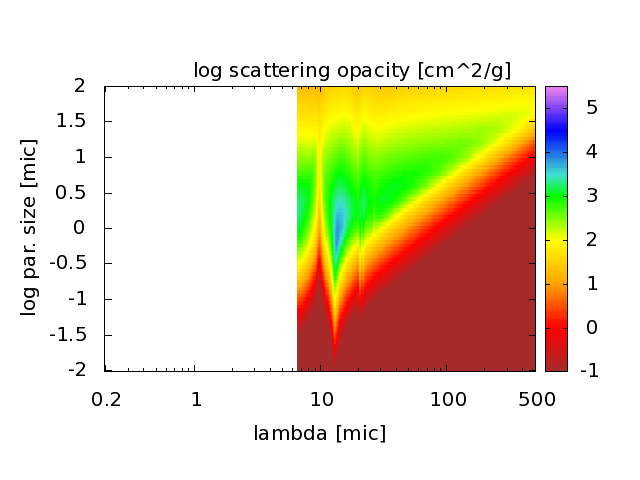}
}
\vspace{-5mm}
\centerline{
\includegraphics[angle=0,width=8.cm]{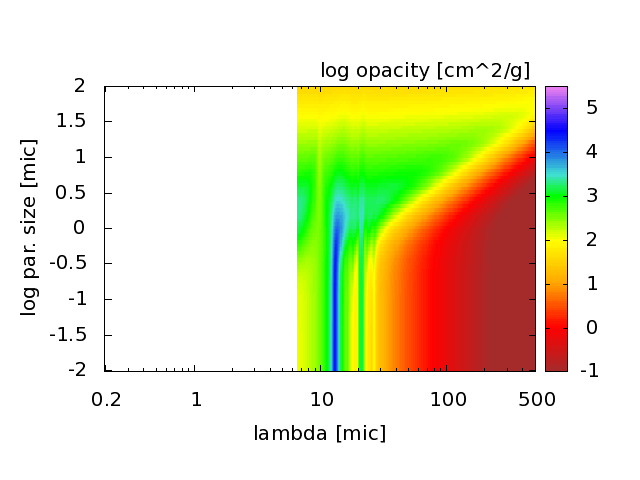}
\includegraphics[angle=0,width=8.cm]{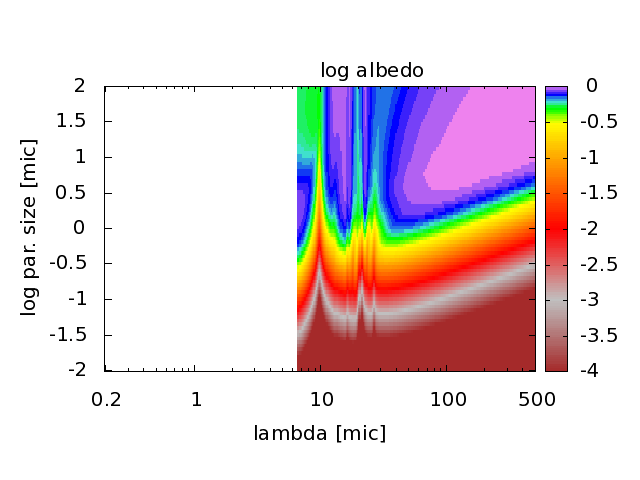}
}
\vspace{-5mm}
\centerline{
\includegraphics[angle=0,width=8.cm]{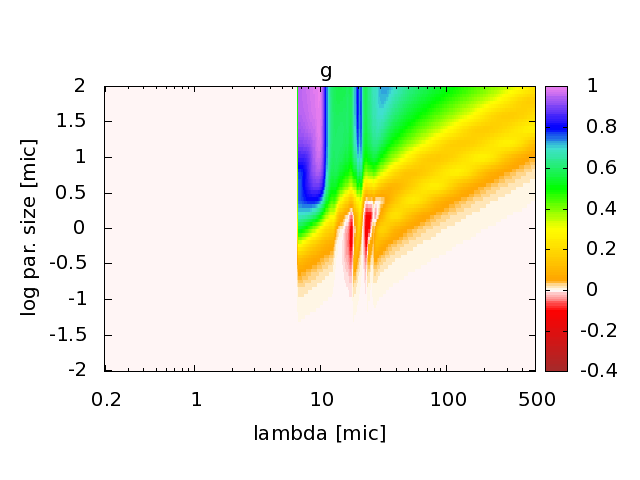}
\includegraphics[angle=0,width=8.cm]{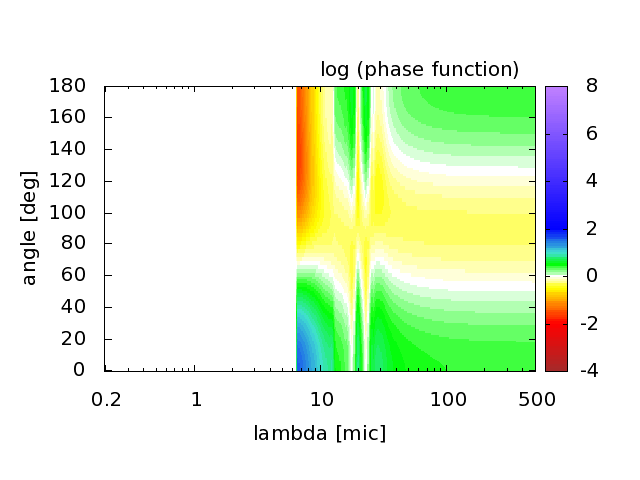}
}
\caption{Corundum. 
Pictures and notation are the same as in the previous figure.
Dust temperatures and accelerations are not calculated
in this case because the opacities do not cover the whole 
spectral region.
}
\label{corundum} 
\end{figure*}

\begin{figure*}
\centerline{
\includegraphics[angle=0,width=8.cm]{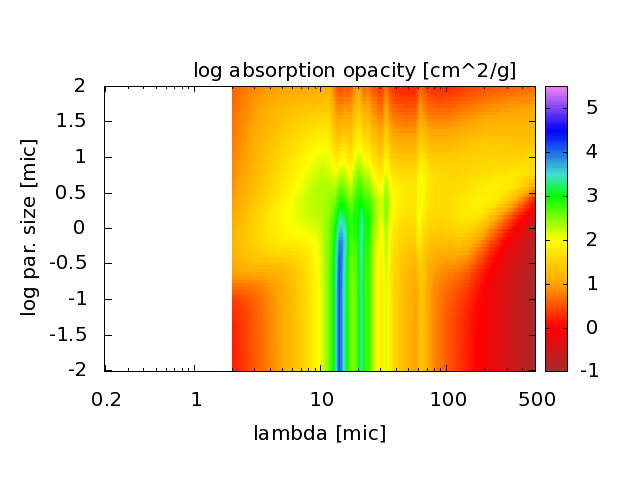}
\includegraphics[angle=0,width=8.cm]{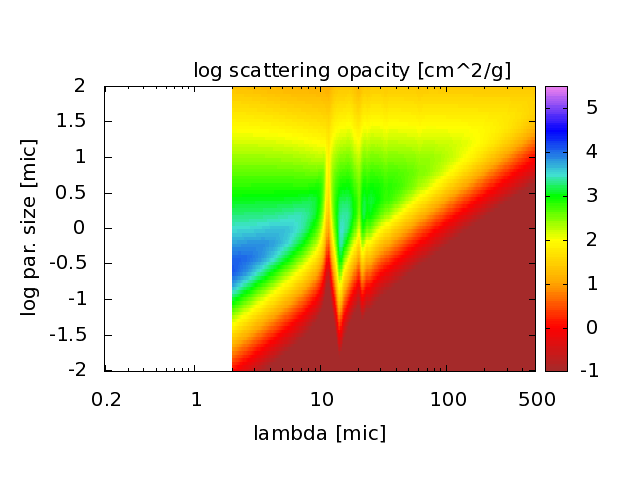}
}
\vspace{-5mm}
\centerline{
\includegraphics[angle=0,width=8.cm]{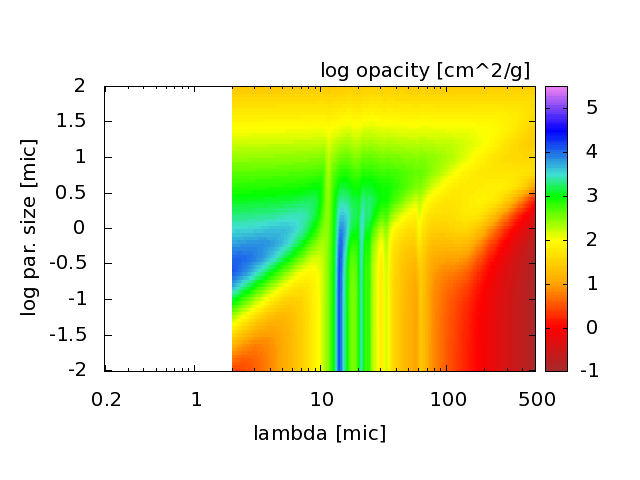}
\includegraphics[angle=0,width=8.cm]{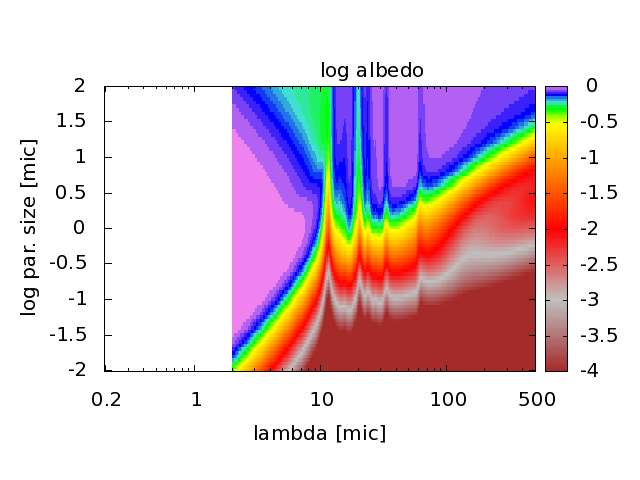}
}
\vspace{-5mm}
\centerline{
\includegraphics[angle=0,width=8.cm]{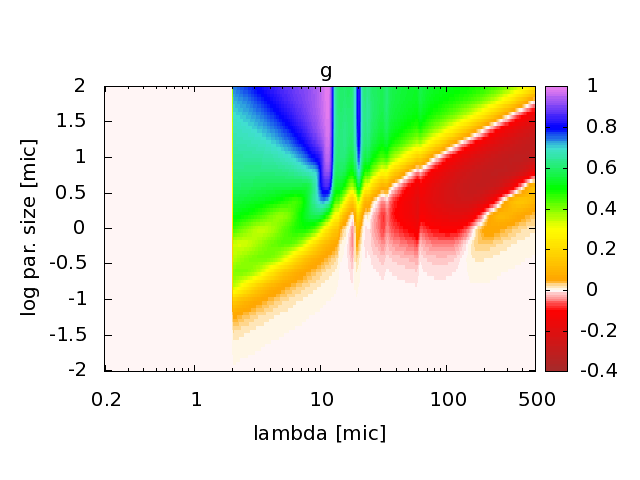}
\includegraphics[angle=0,width=8.cm]{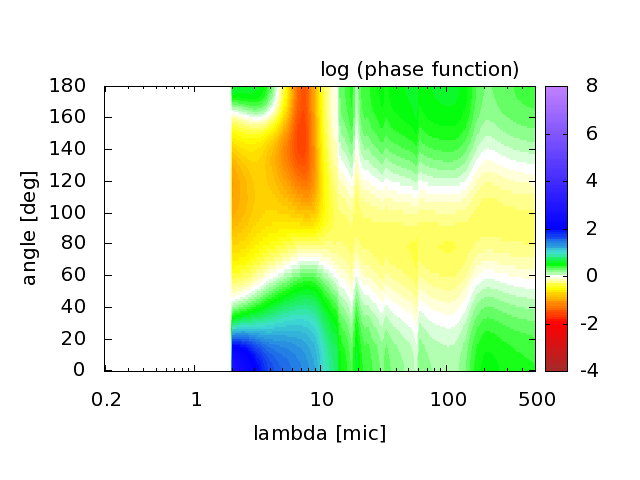}
}
\caption{Perovskite. 
Pictures and notation are the same as in the previous figure.
Dust temperatures and accelerations are not calculated
in this case because the opacities do not cover the whole 
spectral region.
}
\label{perovskite} 
\end{figure*}

\begin{figure*}
\centerline{
\includegraphics[angle=0,width=8.cm]{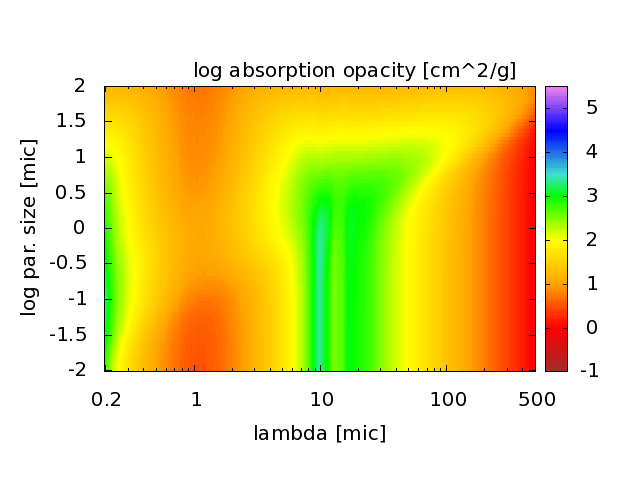}
\includegraphics[angle=0,width=8.cm]{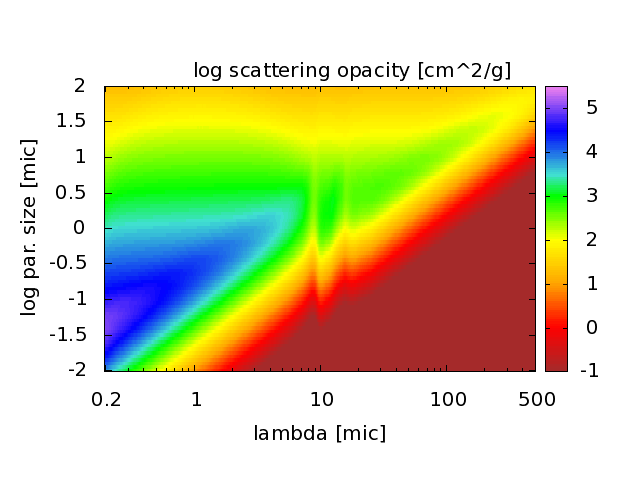}
}
\vspace{-5mm}
\centerline{
\includegraphics[angle=0,width=8.cm]{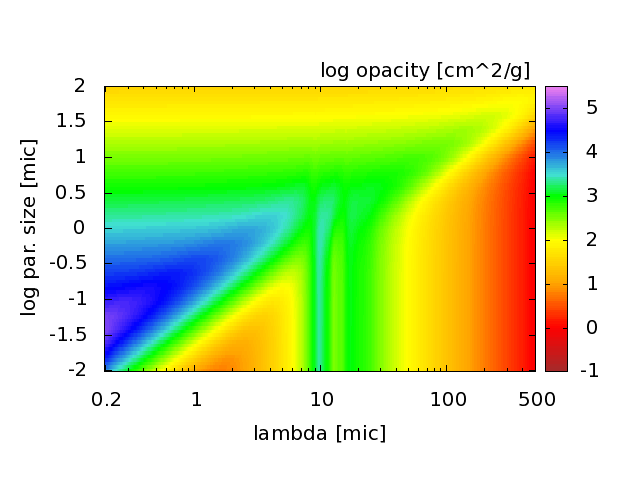}
\includegraphics[angle=0,width=8.cm]{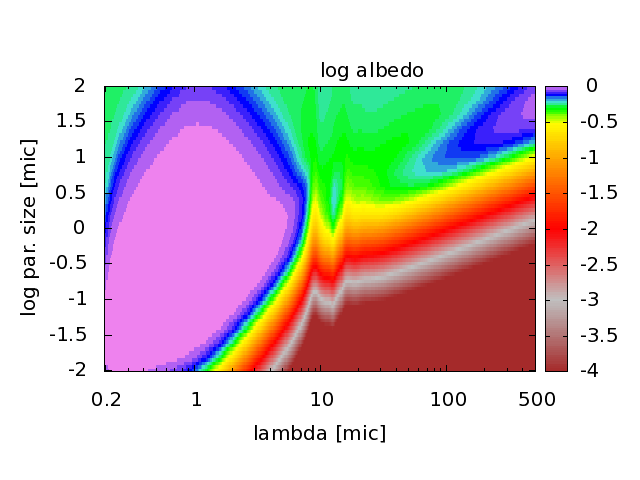}
}
\vspace{-5mm}
\centerline{
\includegraphics[angle=0,width=8.cm]{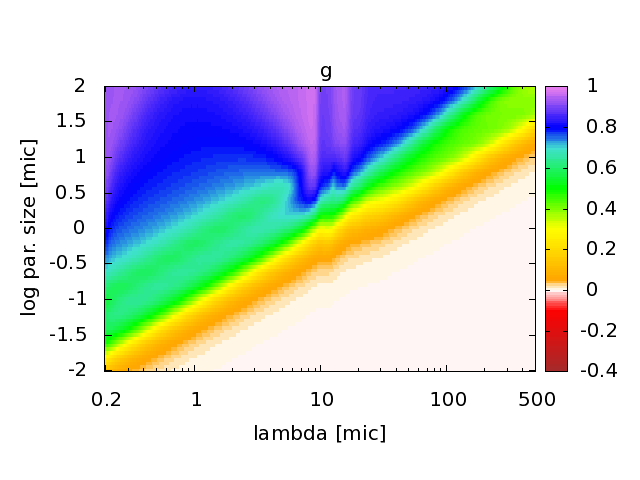}
\includegraphics[angle=0,width=8.cm]{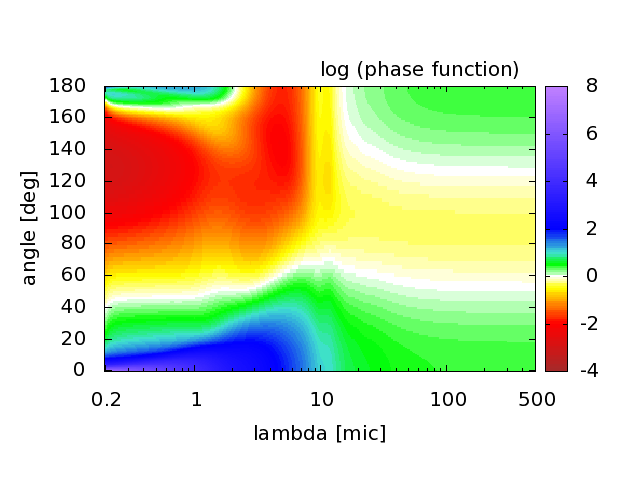}
}
\vspace{-5mm}
\centerline{
\includegraphics[angle=0,width=8.cm]{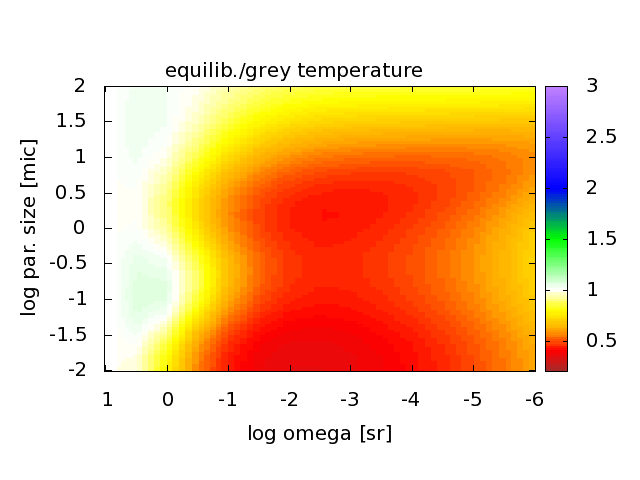}
\includegraphics[angle=0,width=7.5cm]{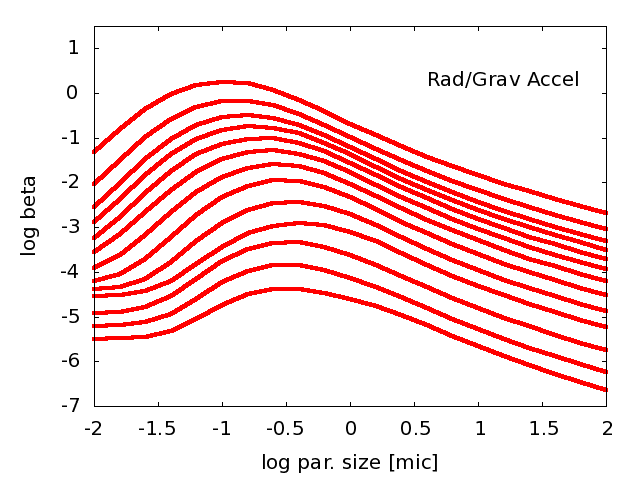}
}
\caption{Forsterite. 
Pictures and notation are the same as in the previous figure.
}
\label{forsterite} 
\end{figure*}

\begin{figure*}
\centerline{
\includegraphics[angle=0,width=8.cm]{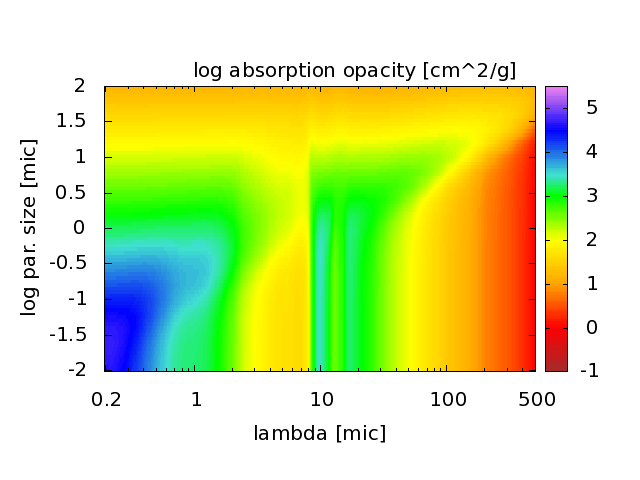}
\includegraphics[angle=0,width=8.cm]{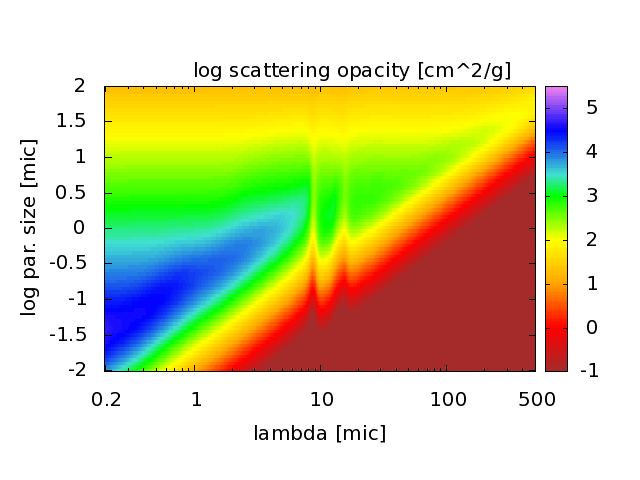}
}
\vspace{-5mm}
\centerline{
\includegraphics[angle=0,width=8.cm]{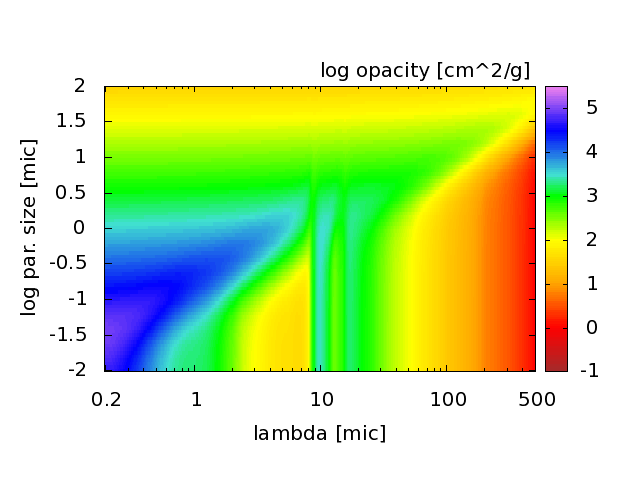}
\includegraphics[angle=0,width=8.cm]{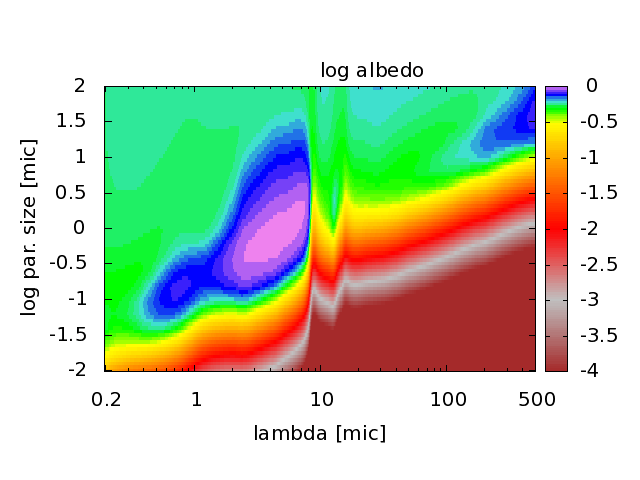}
}
\vspace{-5mm}
\centerline{
\includegraphics[angle=0,width=8.cm]{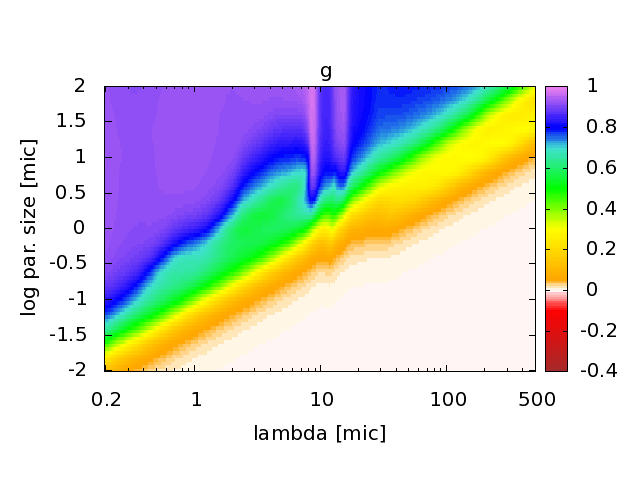}
\includegraphics[angle=0,width=8.cm]{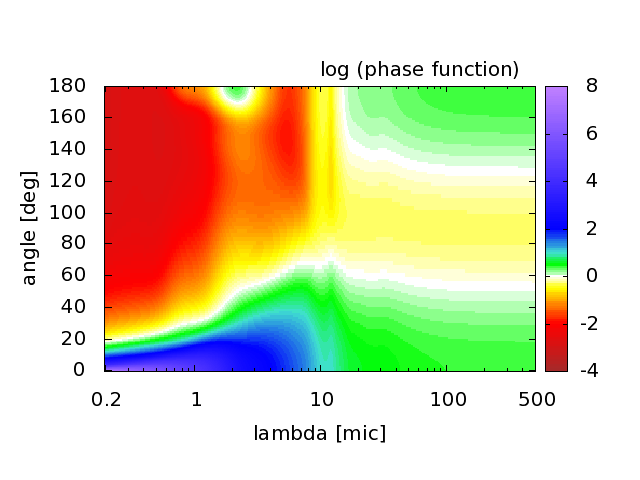}
}
\vspace{-5mm}
\centerline{
\includegraphics[angle=0,width=8.cm]{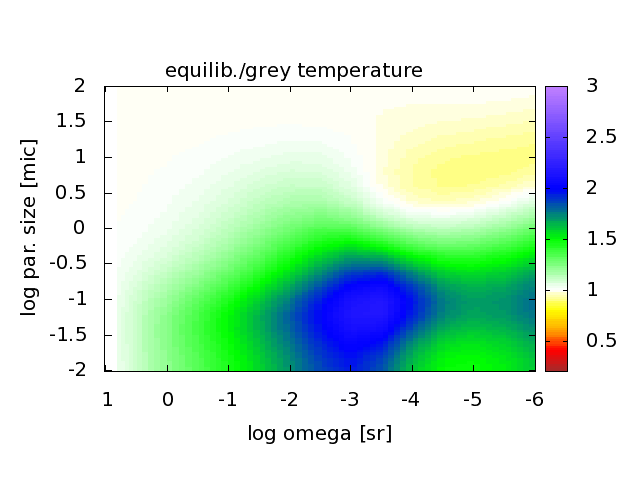}
\includegraphics[angle=0,width=7.5cm]{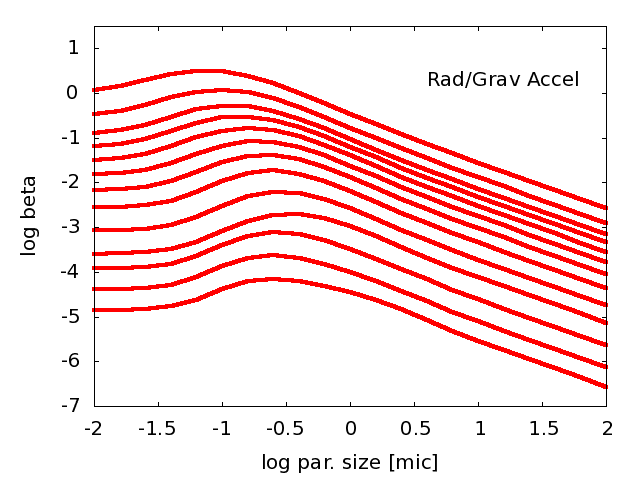}
}
\caption{Olivine with 50\% of iron. 
Pictures and notation are the same as in the previous figure.
}
\label{olivine} 
\end{figure*}

\begin{figure*}
\centerline{   
\includegraphics[angle=0,width=8.cm]{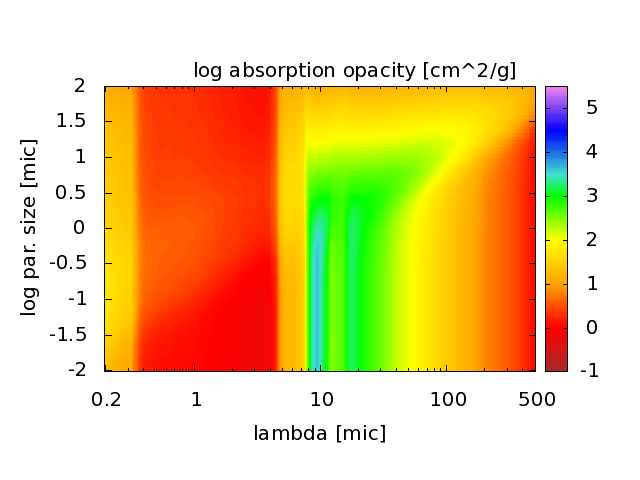}
\includegraphics[angle=0,width=8.cm]{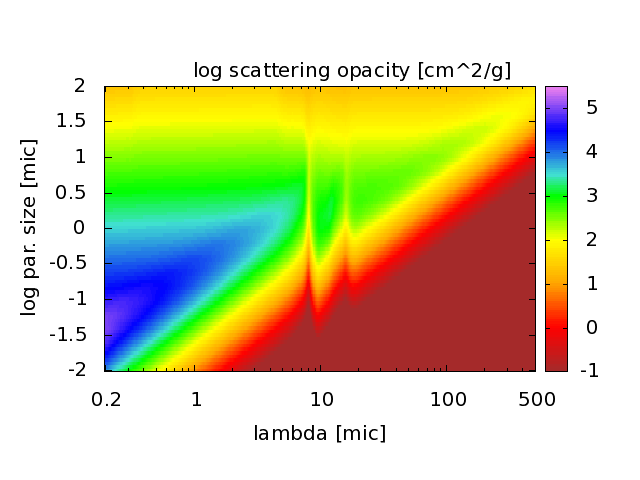}
}
\vspace{-5mm}
\centerline{   
\includegraphics[angle=0,width=8.cm]{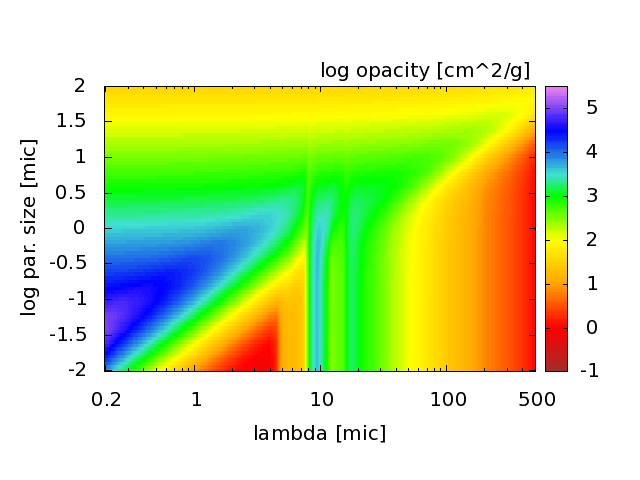}
\includegraphics[angle=0,width=8.cm]{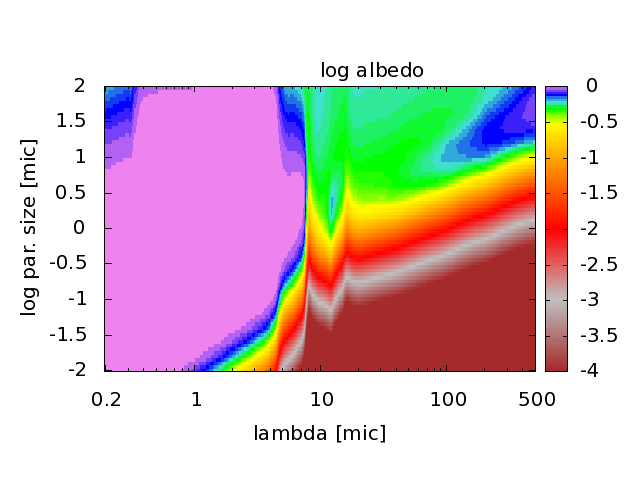}
}
\vspace{-5mm}
\centerline{
\includegraphics[angle=0,width=8.cm]{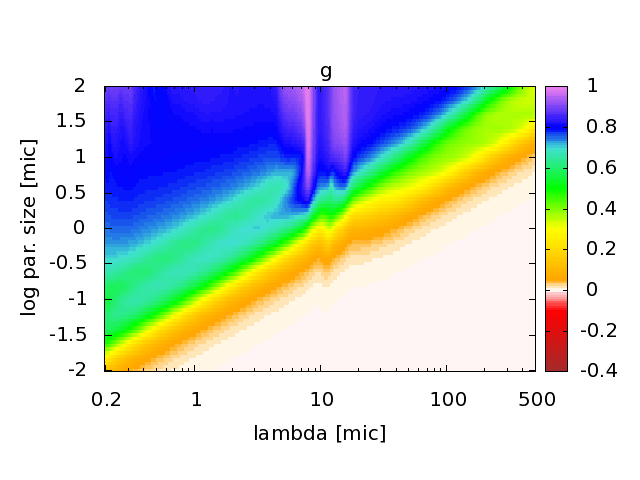}
\includegraphics[angle=0,width=8.cm]{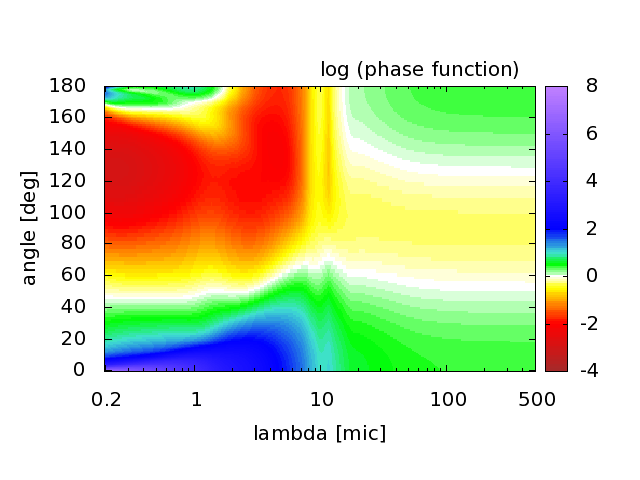}
}
\vspace{-5mm}
\centerline{
\includegraphics[angle=0,width=8.cm]{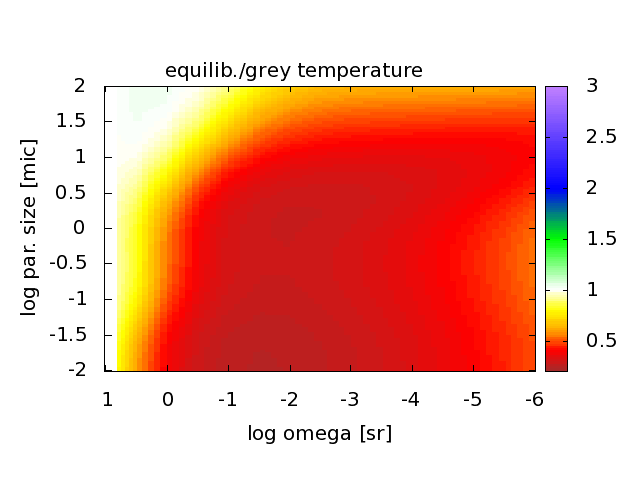}
\includegraphics[angle=0,width=7.5cm]{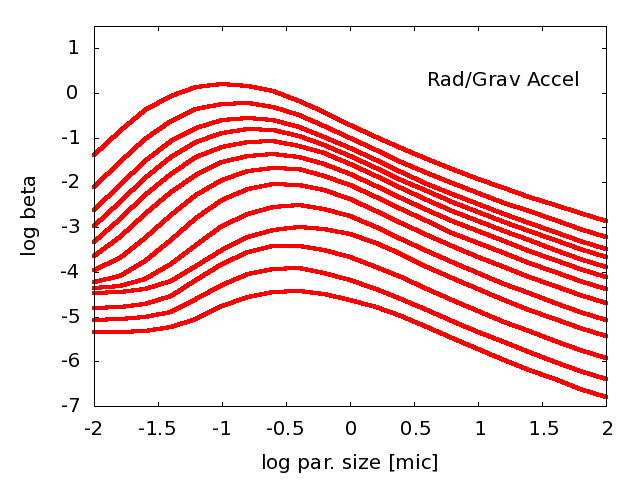}
}
\caption{Enstatite.
Pictures and notation are the same as in the previous figure.
}
\label{enstatite} 
\end{figure*}

\begin{figure*}
\centerline{   
\includegraphics[angle=0,width=8.cm]{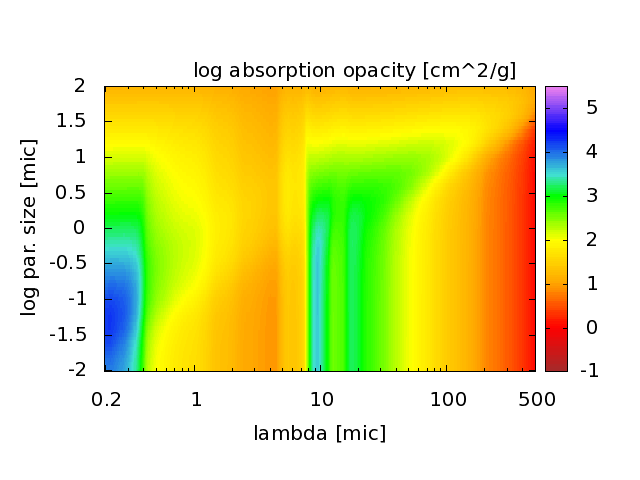}
\includegraphics[angle=0,width=8.cm]{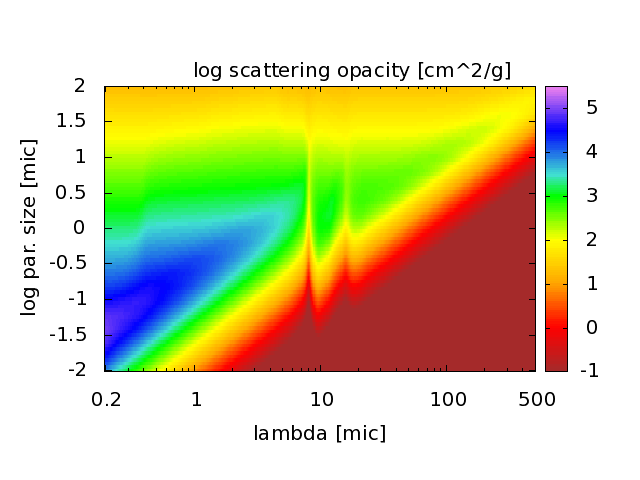}
}
\vspace{-5mm}
\centerline{   
\includegraphics[angle=0,width=8.cm]{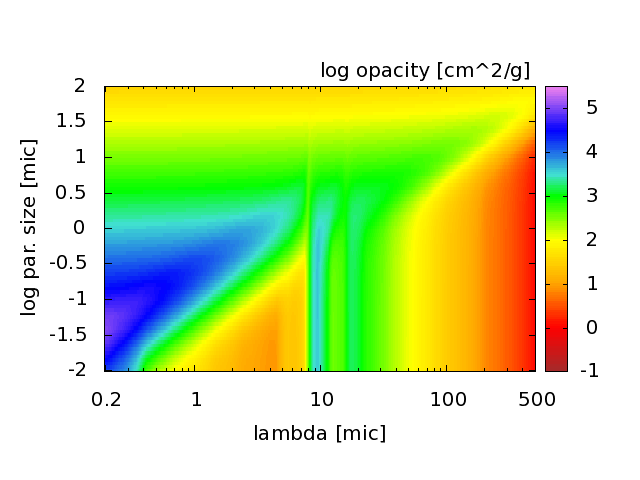}
\includegraphics[angle=0,width=8.cm]{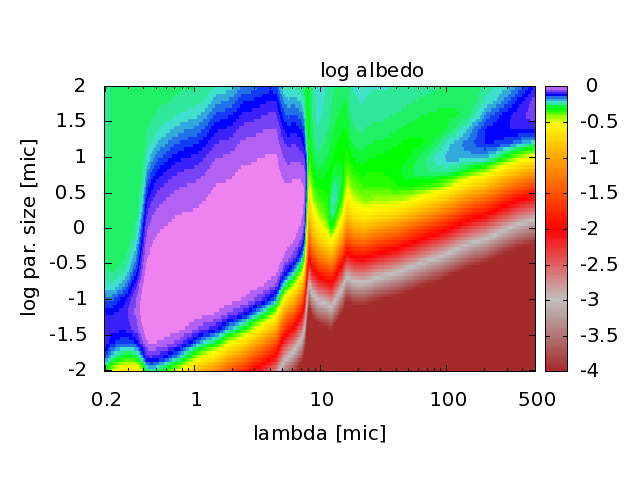}
}
\vspace{-5mm}
\centerline{
\includegraphics[angle=0,width=8.cm]{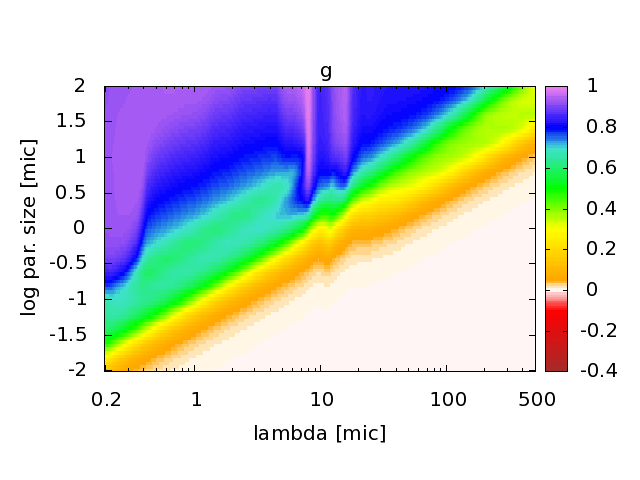}
\includegraphics[angle=0,width=8.cm]{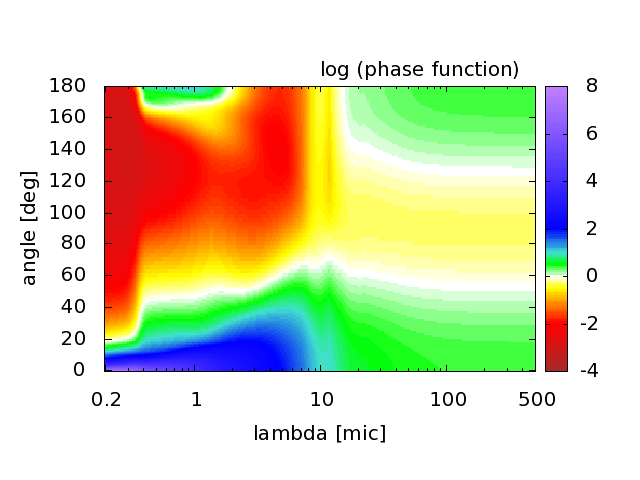}
}
\vspace{-5mm}
\centerline{
\includegraphics[angle=0,width=8.cm]{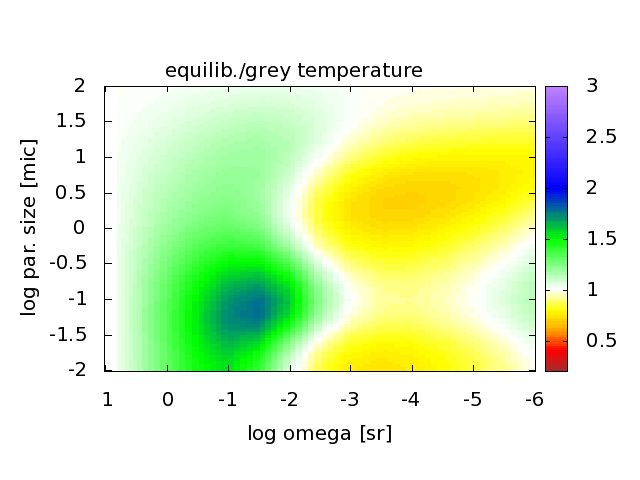}
\includegraphics[angle=0,width=7.5cm]{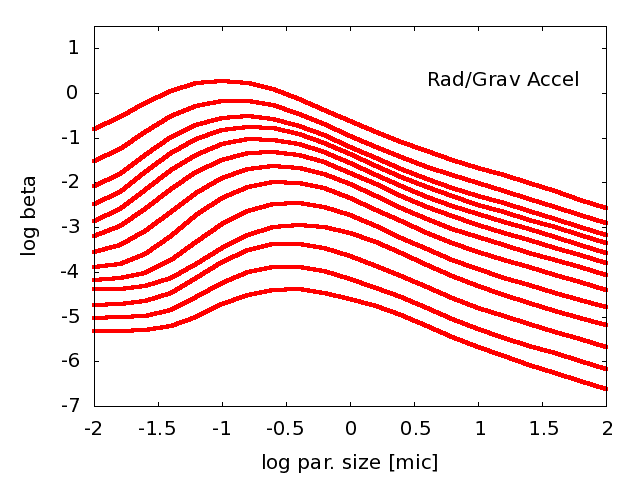}
}
\caption{Pyroxene with 20\% or iron.
Pictures and notation are the same as in the previous figure.
}
\label{pyrmg80}
\end{figure*}

\begin{figure*}
\centerline{   
\includegraphics[angle=0,width=8.cm]{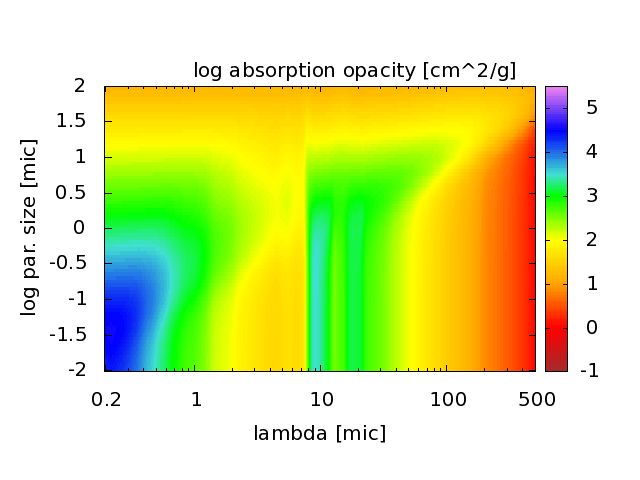}
\includegraphics[angle=0,width=8.cm]{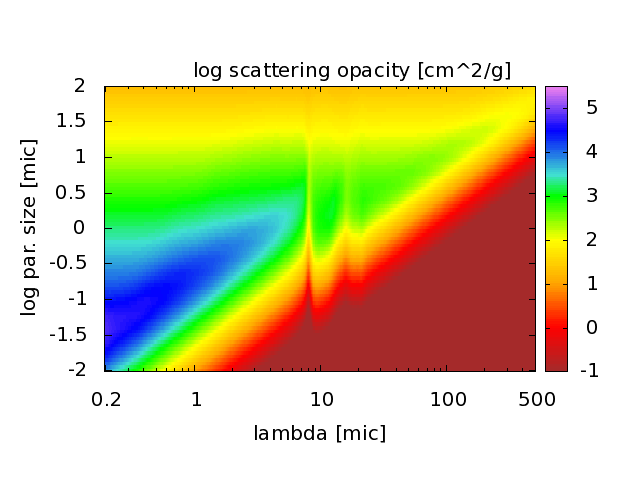}        
}
\vspace{-5mm}
\centerline{   
\includegraphics[angle=0,width=8.cm]{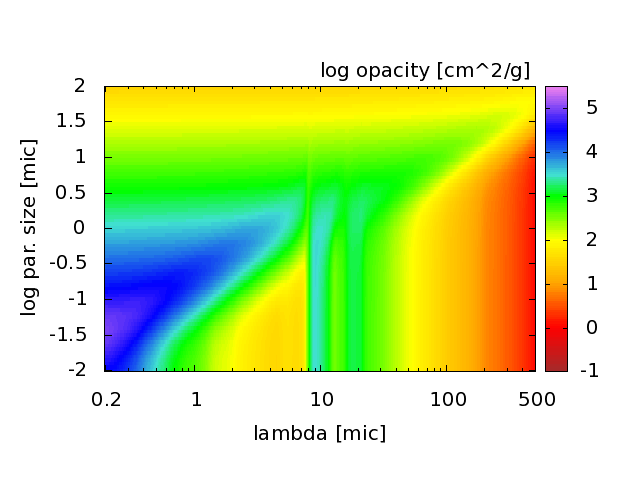}
\includegraphics[angle=0,width=8.cm]{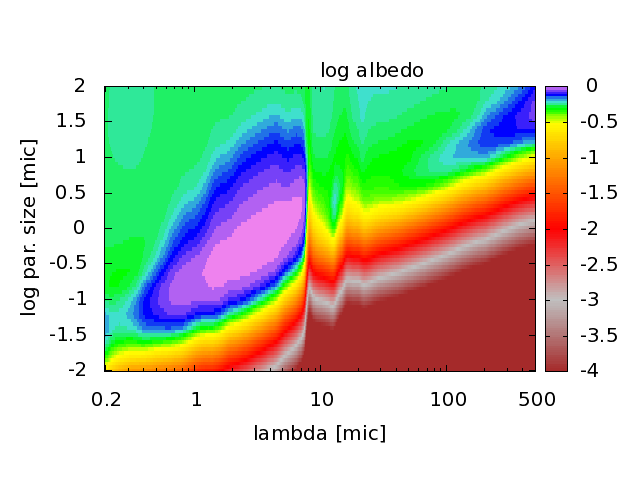}        
}
\vspace{-5mm}
\centerline{
\includegraphics[angle=0,width=8.cm]{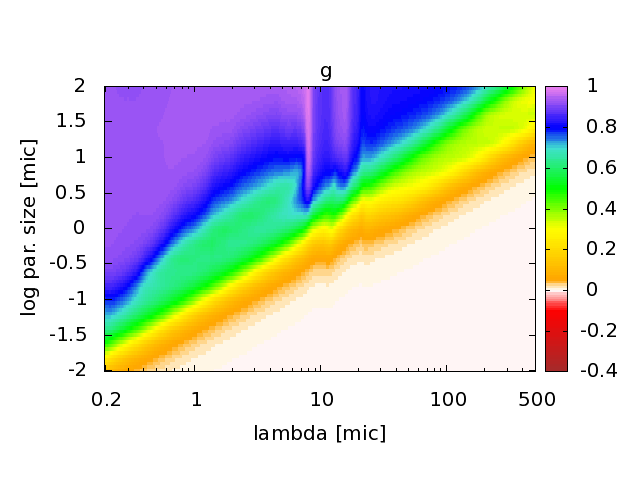}
\includegraphics[angle=0,width=8.cm]{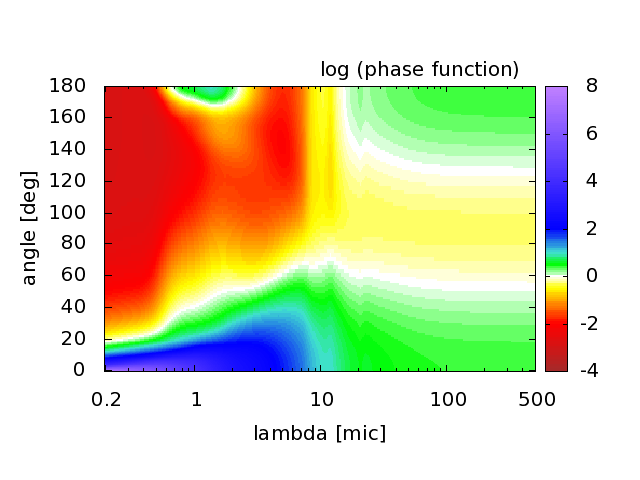}
}
\vspace{-5mm}
\centerline{   
\includegraphics[angle=0,width=8.cm]{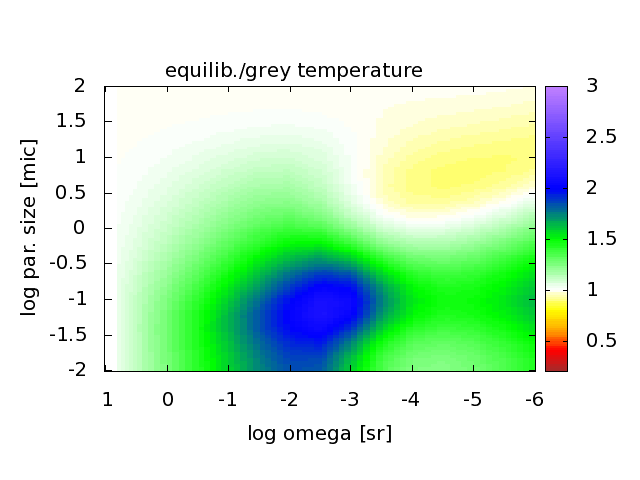}  
\includegraphics[angle=0,width=7.5cm]{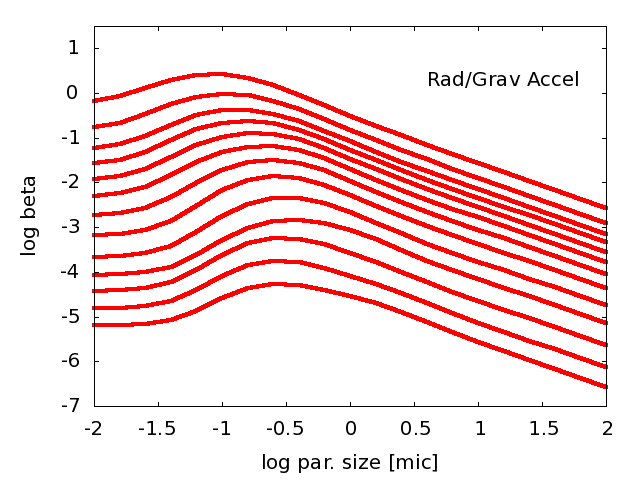}      
}
\caption{Pyroxene with 60\% or iron.
Pictures and notation are the same as in the previous figure.
}
\label{pyrmg40}      
\end{figure*} 

\begin{figure*}
\centerline{
\includegraphics[angle=0,width=8.cm]{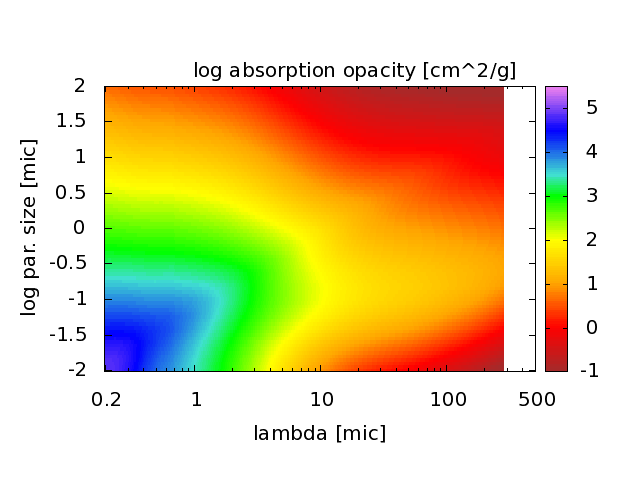}
\includegraphics[angle=0,width=8.cm]{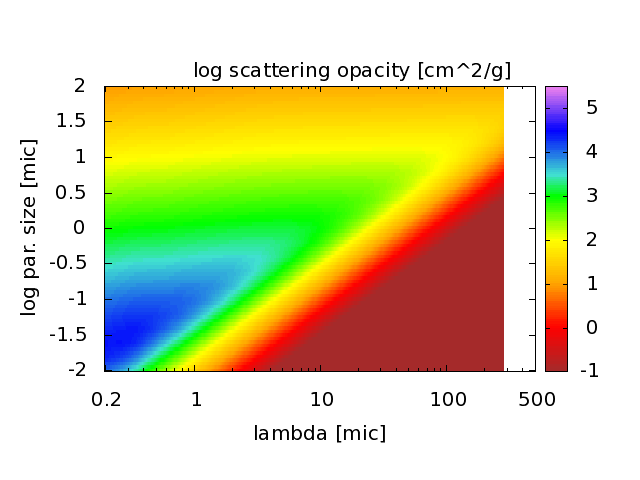}
}
\vspace{-5mm}
\centerline{
\includegraphics[angle=0,width=8.cm]{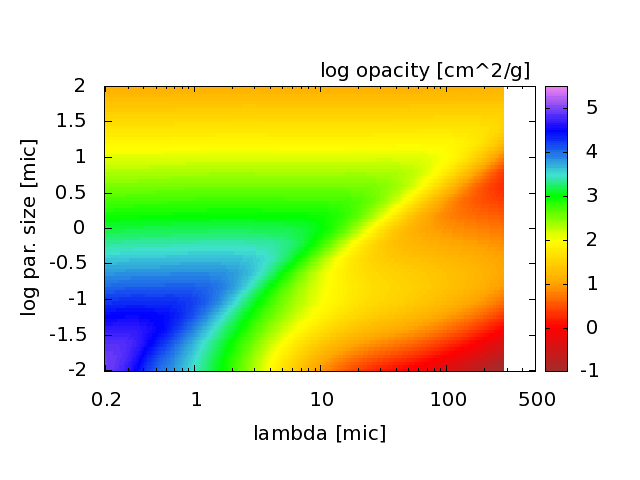}
\includegraphics[angle=0,width=8.cm]{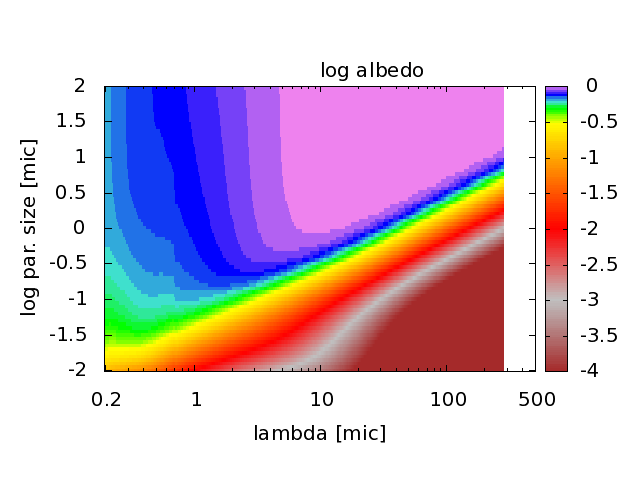}
}
\vspace{-5mm}
\centerline{
\includegraphics[angle=0,width=8.cm]{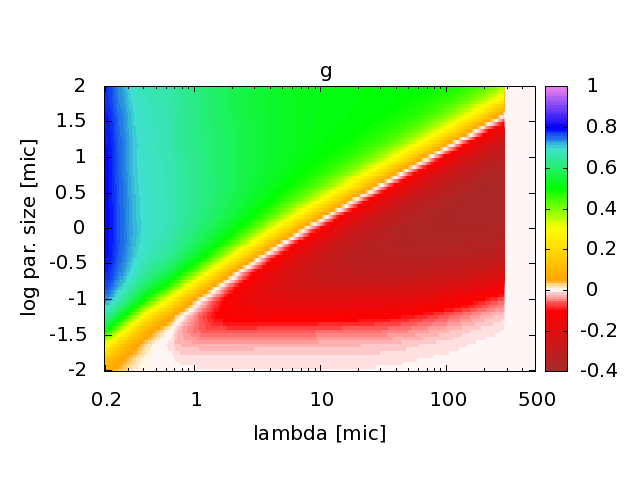}
\includegraphics[angle=0,width=8.cm]{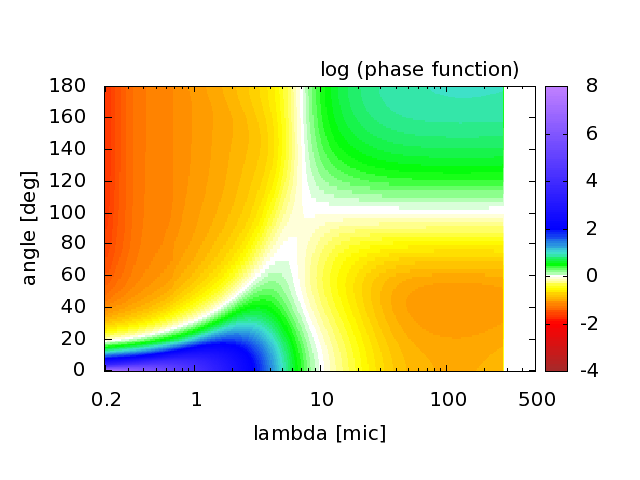}
}
\vspace{-5mm}
\centerline{
\includegraphics[angle=0,width=8.cm]{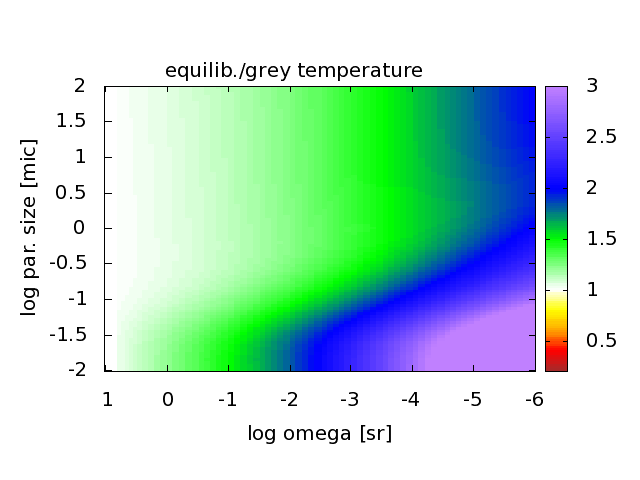}
\includegraphics[angle=0,width=7.5cm]{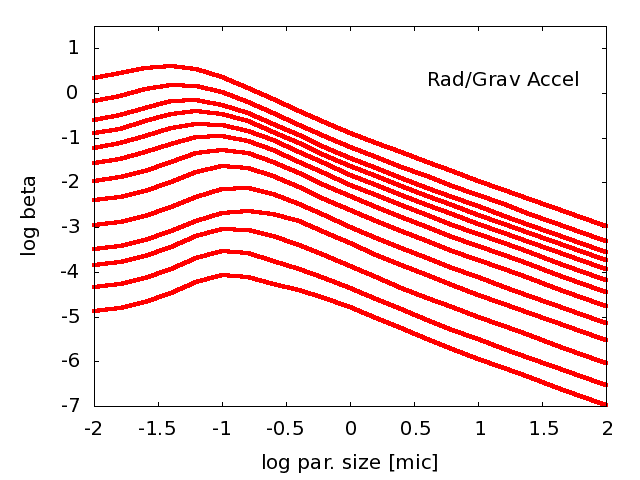}
}
\caption{Iron. 
Pictures and notation are the same as in the previous figure.
}
\label{iron} 
\end{figure*}

\begin{figure*}
\centerline{
\includegraphics[angle=0,width=8.cm]{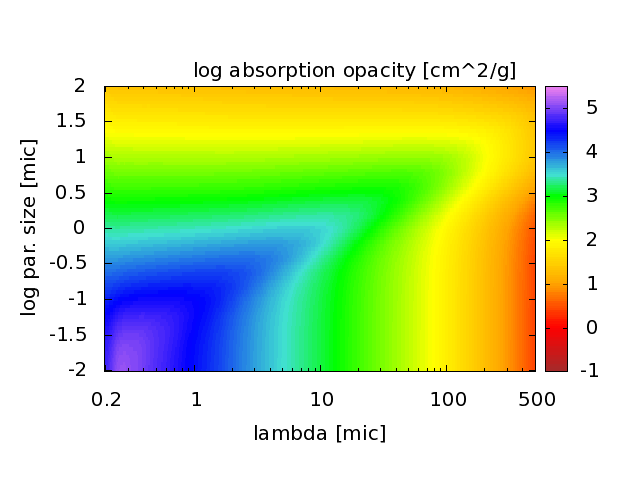}
\includegraphics[angle=0,width=8.cm]{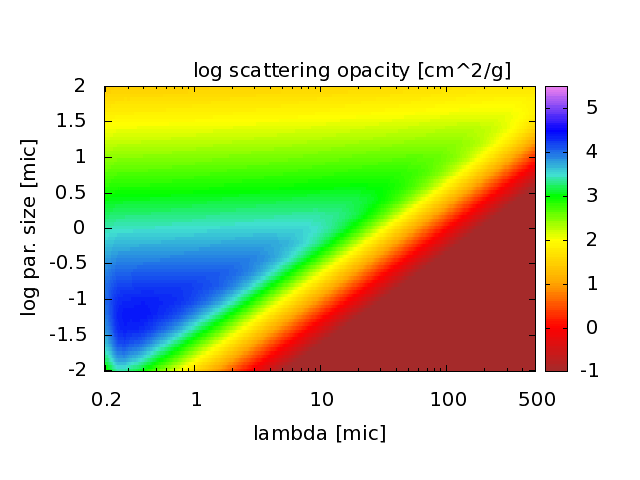}
}
\vspace{-5mm}
\centerline{
\includegraphics[angle=0,width=8.cm]{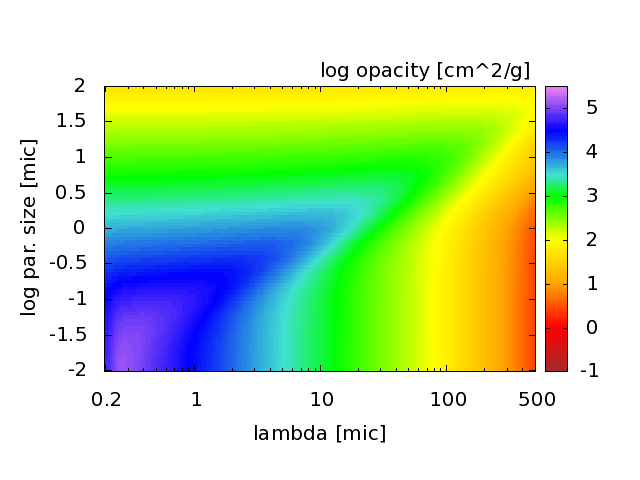}
\includegraphics[angle=0,width=8.cm]{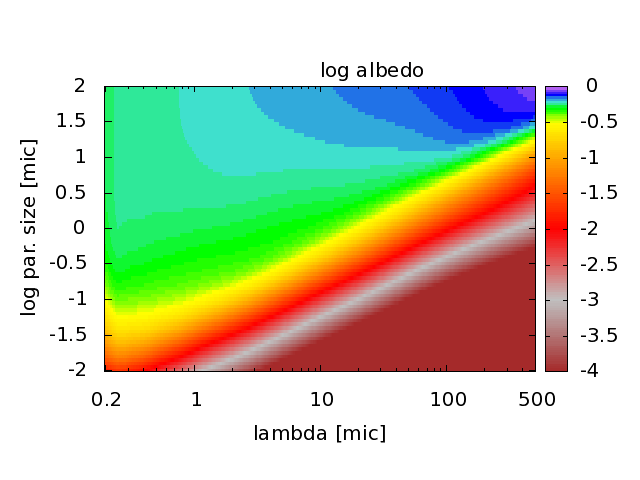}
}
\vspace{-5mm}
\centerline{
\includegraphics[angle=0,width=8.cm]{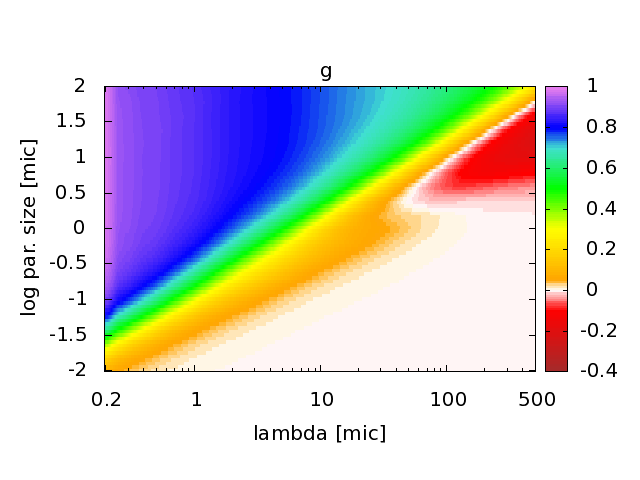}
\includegraphics[angle=0,width=8.cm]{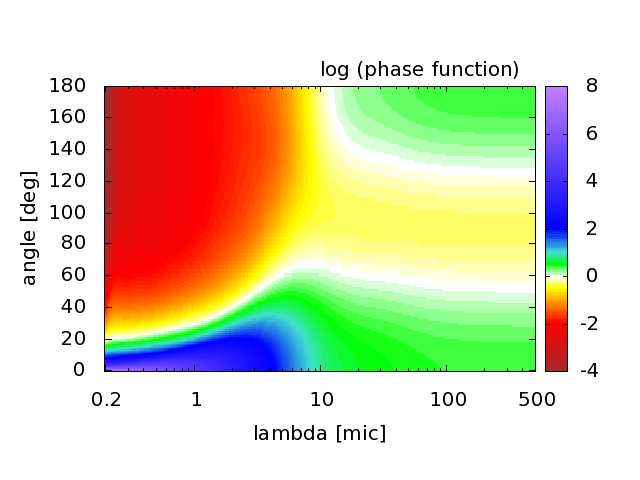}
}
\vspace{-5mm}
\centerline{
\includegraphics[angle=0,width=8.cm]{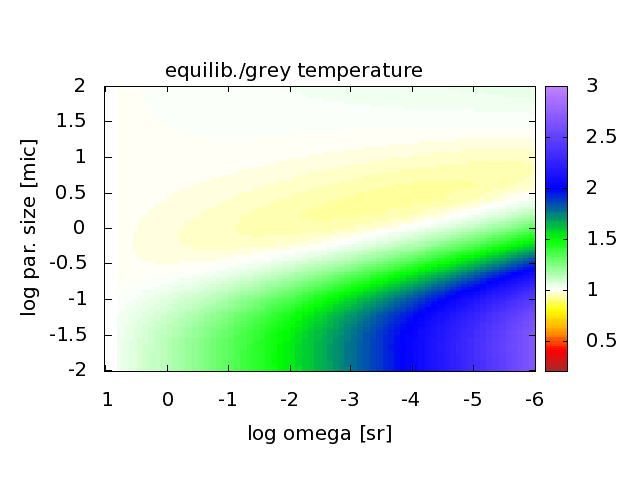}
\includegraphics[angle=0,width=7.5cm]{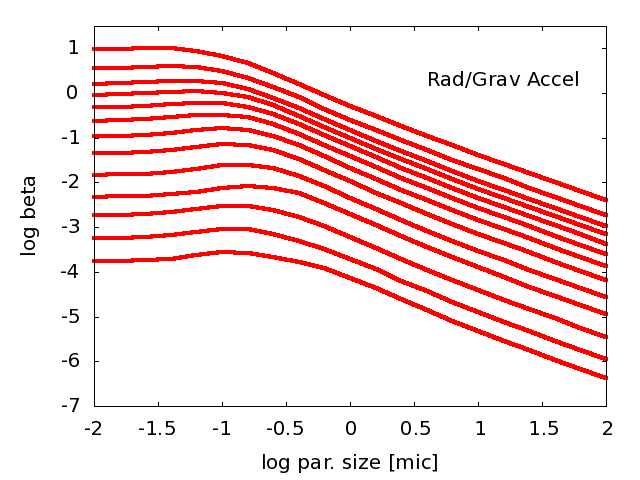}
}
\caption{Carbon assuming optical data for 1000 C. 
Pictures and notation are the same as in the previous figure.
}
\label{carbon1000} 
\end{figure*}

\begin{figure*}
\centerline{
\includegraphics[angle=0,width=8.cm]{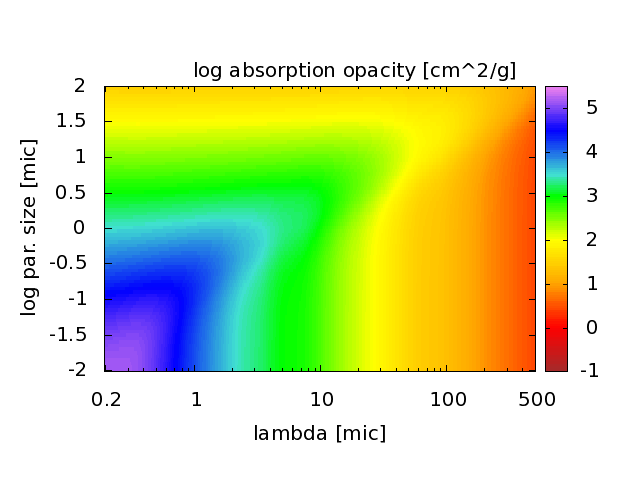}
\includegraphics[angle=0,width=8.cm]{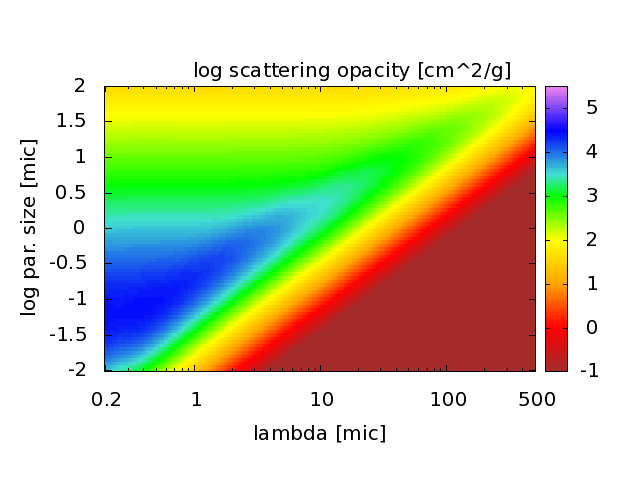}
}
\vspace{-5mm}
\centerline{
\includegraphics[angle=0,width=8.cm]{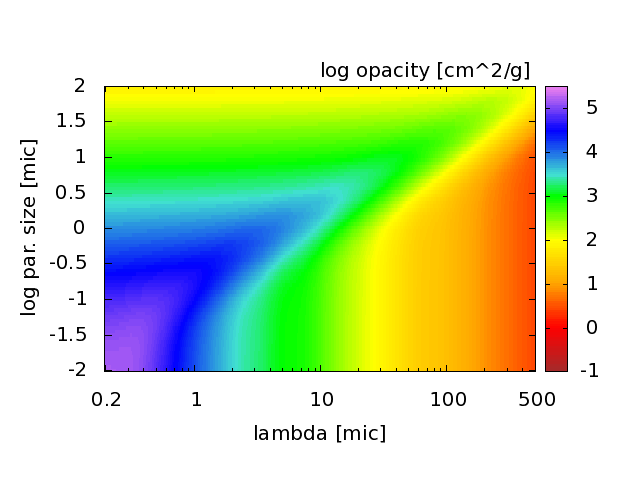}
\includegraphics[angle=0,width=8.cm]{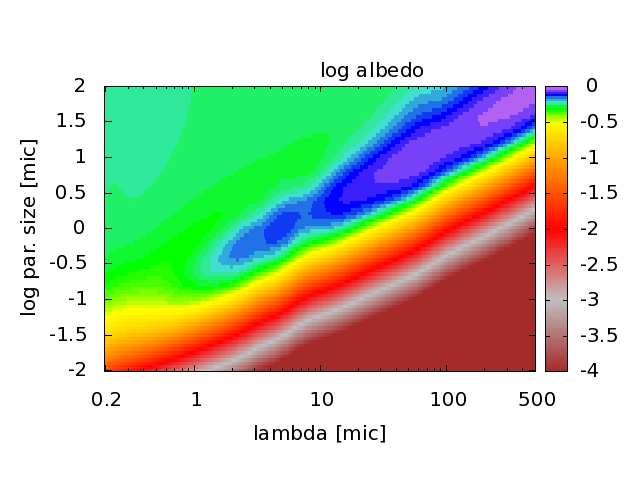}
}
\vspace{-5mm}
\centerline{
\includegraphics[angle=0,width=8.cm]{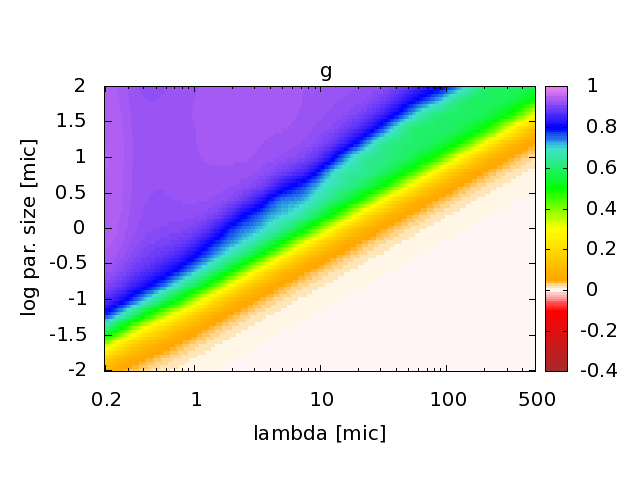}
\includegraphics[angle=0,width=8.cm]{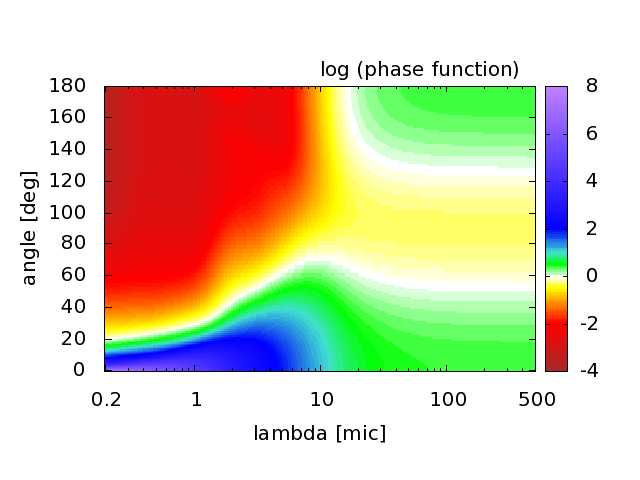}
}
\vspace{-5mm}
\centerline{
\includegraphics[angle=0,width=8.cm]{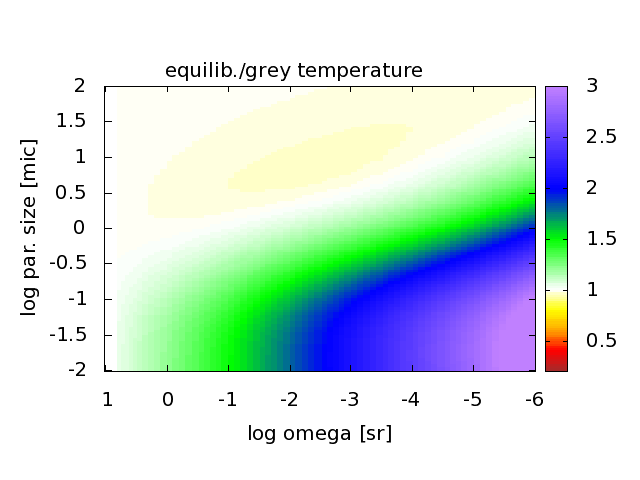}
\includegraphics[angle=0,width=7.5cm]{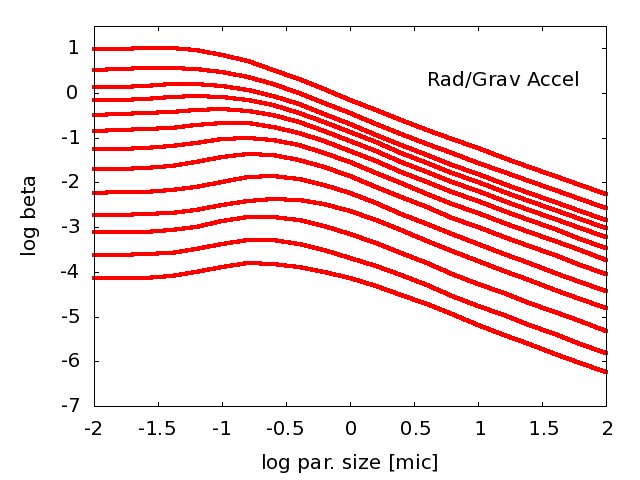}
}
\caption{Carbon assuming optical data for 400 C. 
Pictures and notation are the same as in the previous figure.
}
\label{carbon0400} 
\end{figure*}

\begin{figure*}
\centerline{   
\includegraphics[angle=0,width=8.cm]{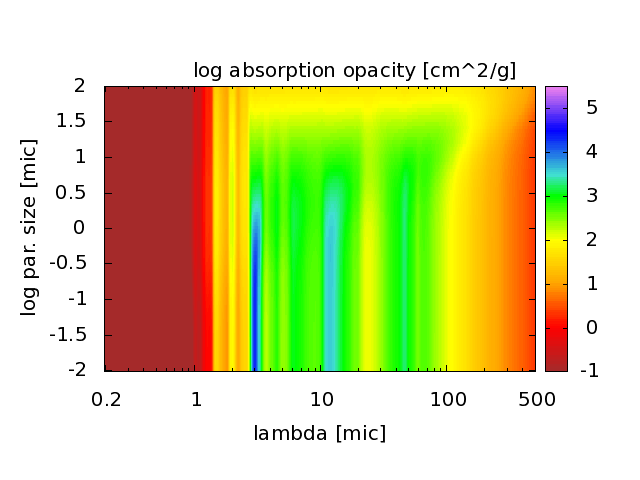}
\includegraphics[angle=0,width=8.cm]{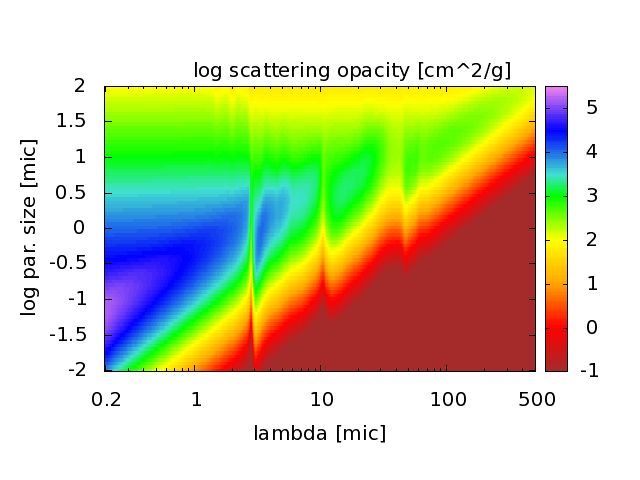}
}
\vspace{-5mm}
\centerline{   
\includegraphics[angle=0,width=8.cm]{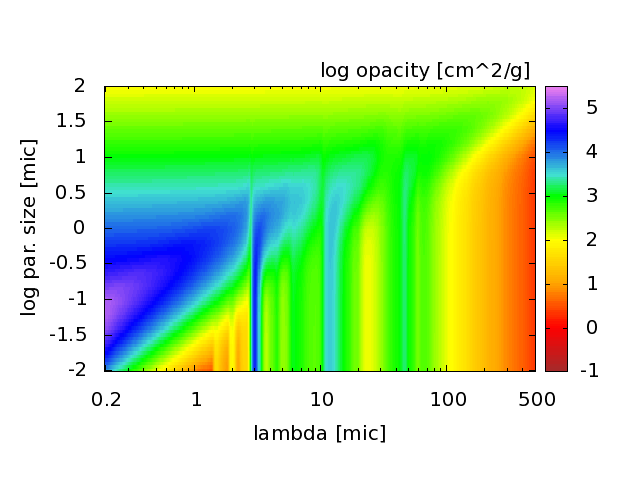}
\includegraphics[angle=0,width=8.cm]{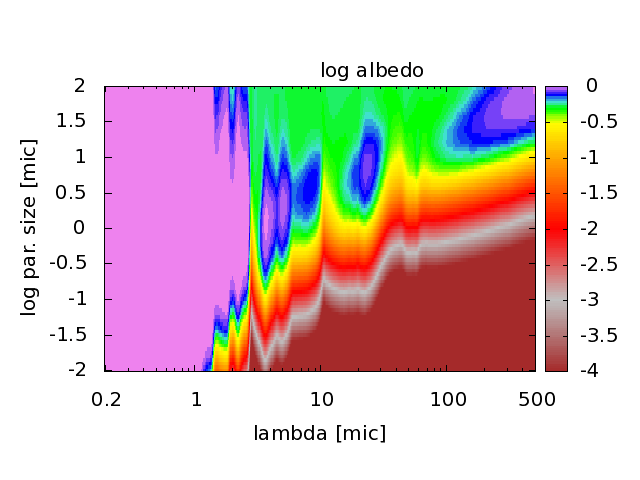}
}
\vspace{-5mm}
\centerline{
\includegraphics[angle=0,width=8.cm]{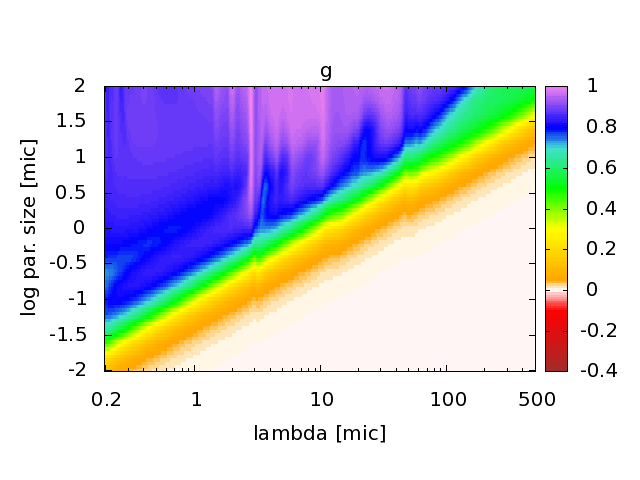}
\includegraphics[angle=0,width=8.cm]{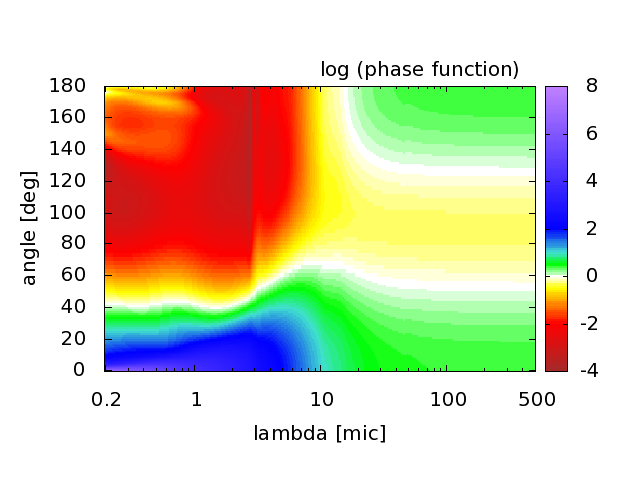}
}
\vspace{-5mm}
\centerline{
\includegraphics[angle=0,width=8.cm]{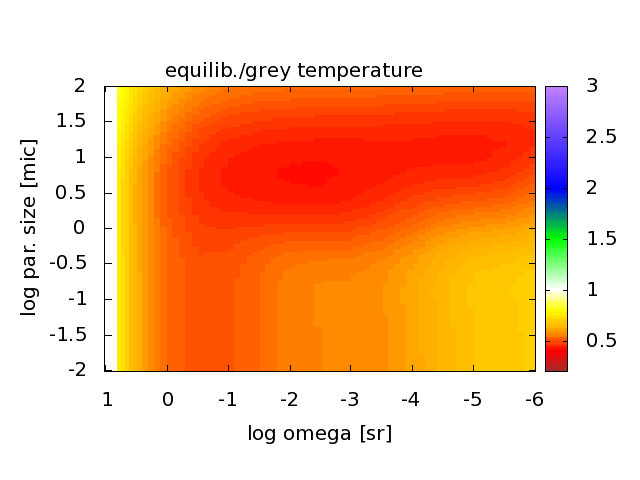}
\includegraphics[angle=0,width=7.5cm]{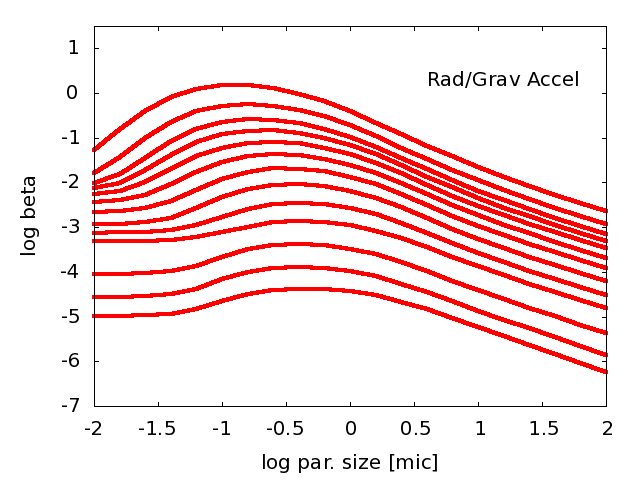}      
}
\caption{Water ice.
Pictures and notation are the same as in the previous figure.
}
\label{waterice}
\end{figure*}

\begin{figure*}
\centerline{   
\includegraphics[angle=0,width=8.cm]{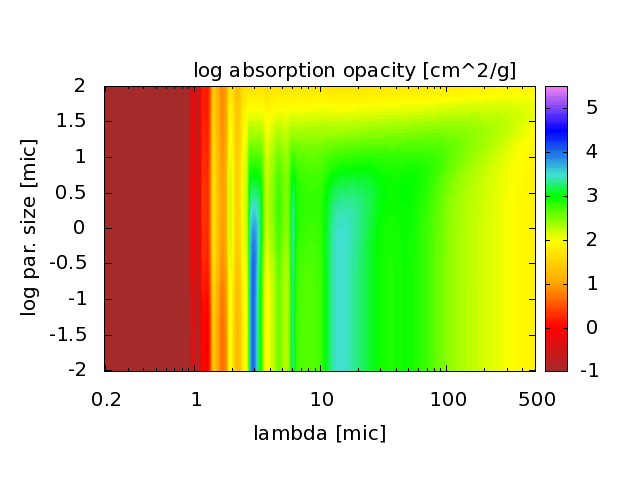} 
\includegraphics[angle=0,width=8.cm]{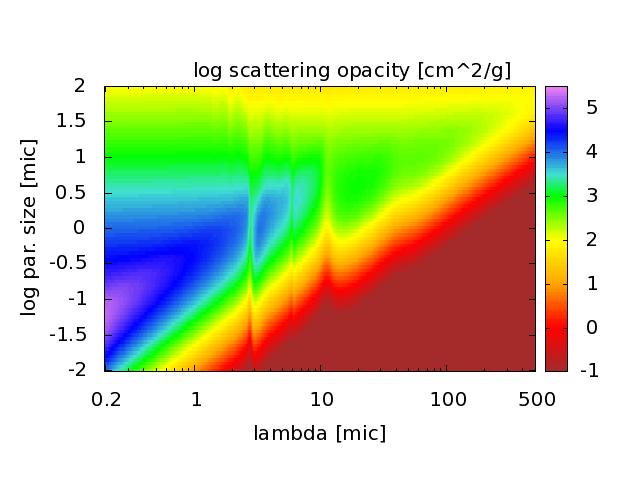}        
}
\vspace{-5mm}
\centerline{   
\includegraphics[angle=0,width=8.cm]{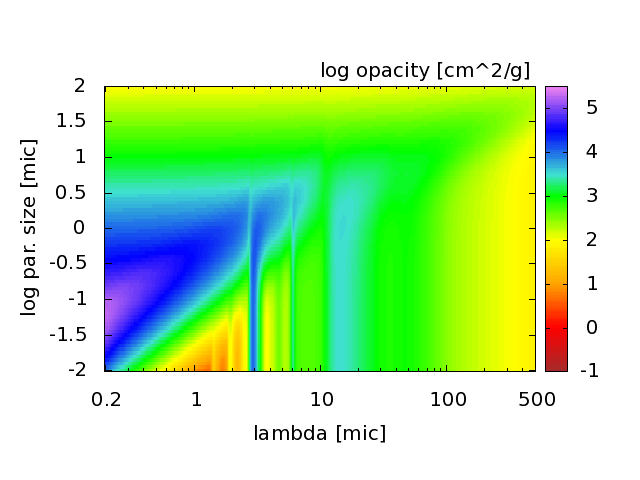} 
\includegraphics[angle=0,width=8.cm]{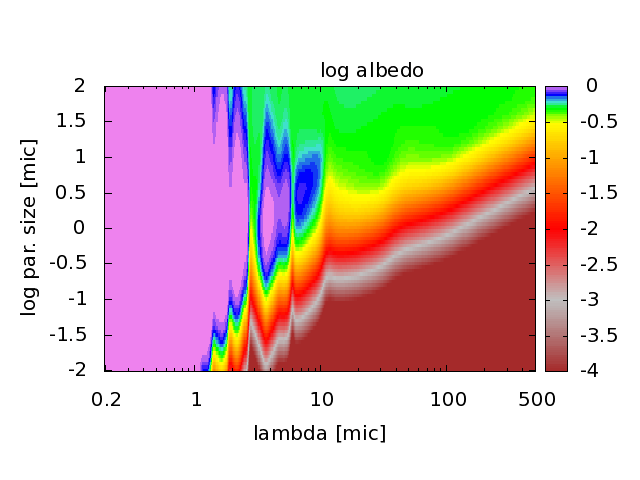}        
}
\vspace{-5mm}
\centerline{
\includegraphics[angle=0,width=8.cm]{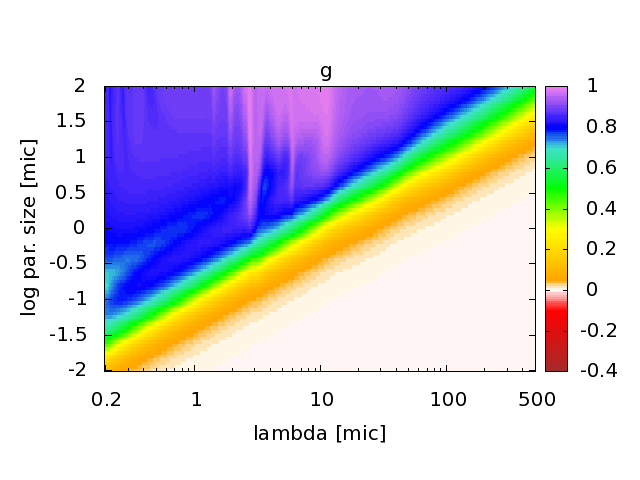}
\includegraphics[angle=0,width=8.cm]{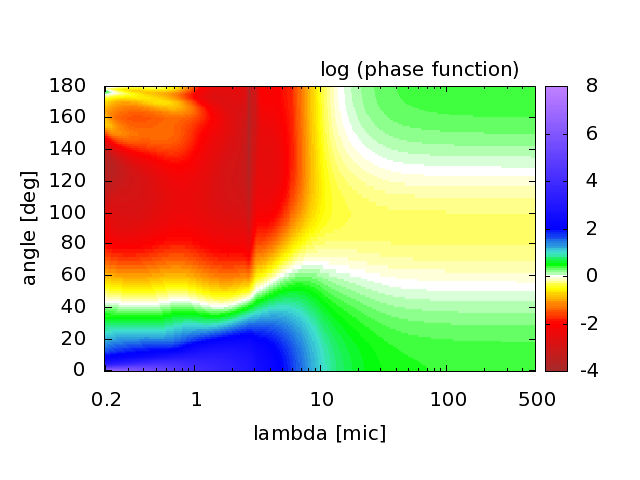}
}
\vspace{-5mm}
\centerline{   
\includegraphics[angle=0,width=8.cm]{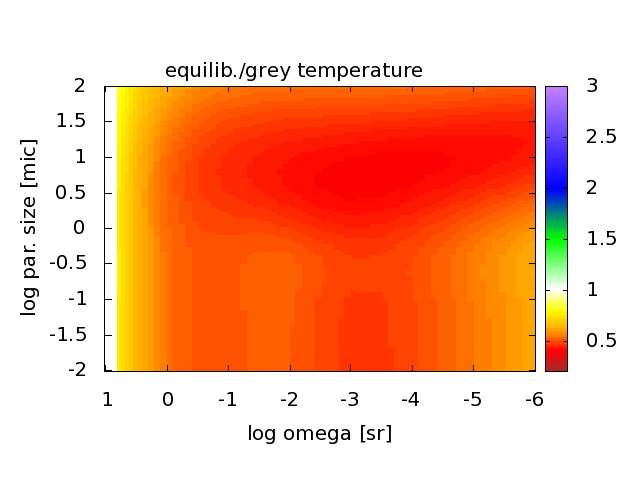}  
\includegraphics[angle=0,width=7.5cm]{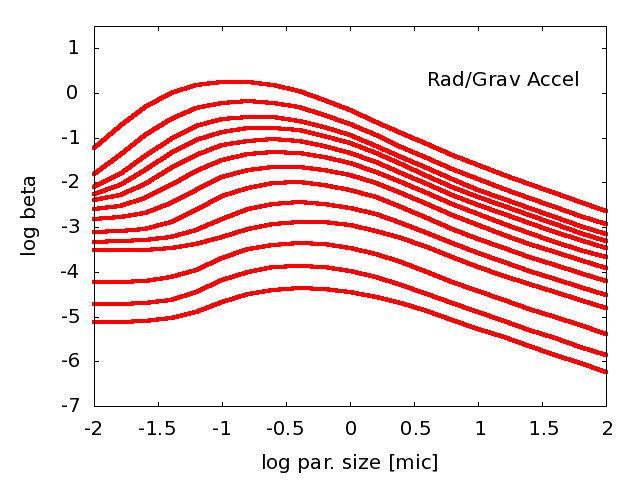}      
}
\caption{Liquid water.
Pictures and notation are the same as in the previous figure.
}
\label{waterliq}      
\end{figure*} 

\begin{figure*}
\centerline{   
\includegraphics[angle=0,width=8.cm]{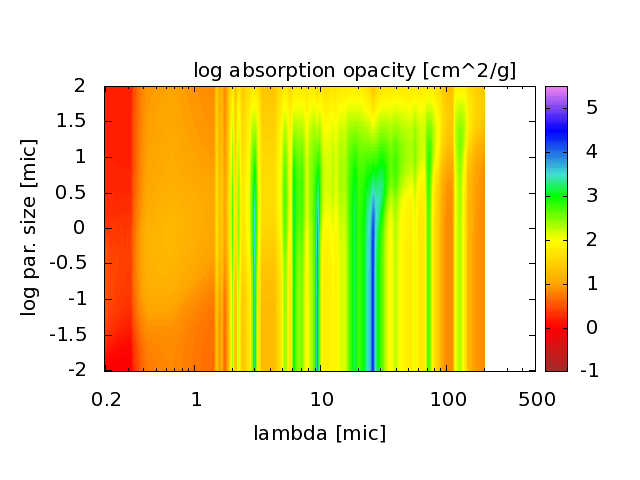} 
\includegraphics[angle=0,width=8.cm]{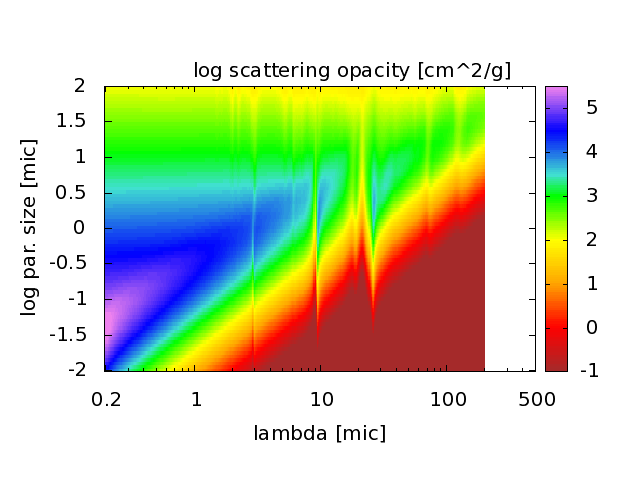}   
}
\vspace{-5mm}
\centerline{   
\includegraphics[angle=0,width=8.cm]{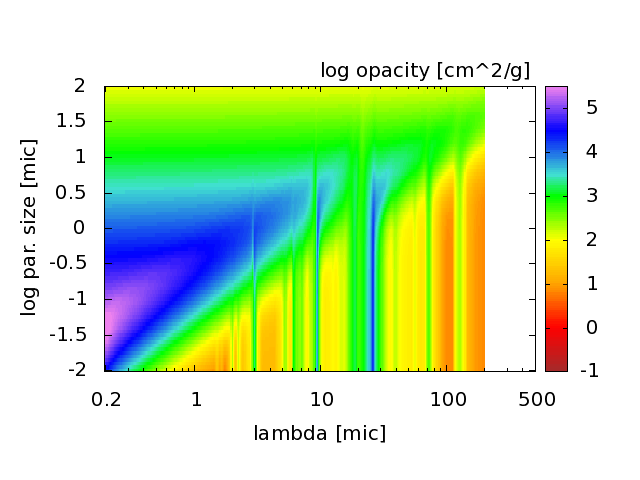} 
\includegraphics[angle=0,width=8.cm]{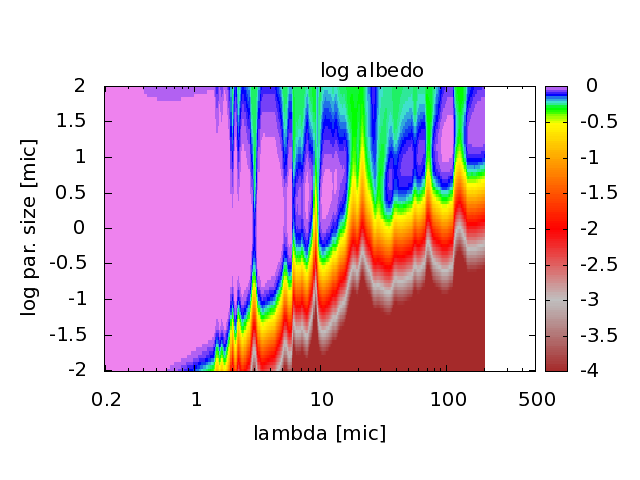}   
}
\vspace{-5mm}
\centerline{
\includegraphics[angle=0,width=8.cm]{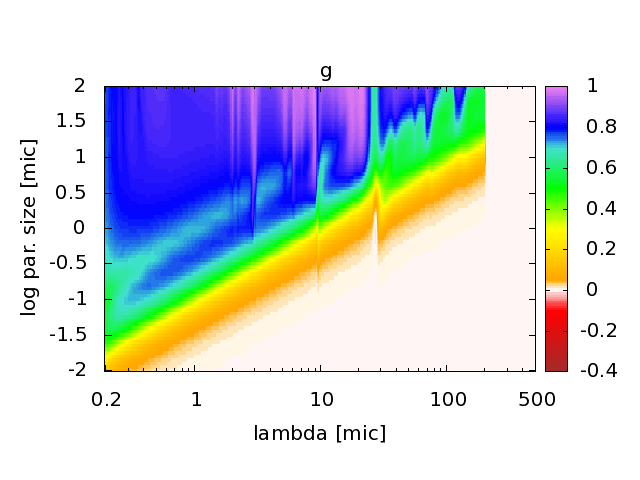}
\includegraphics[angle=0,width=8.cm]{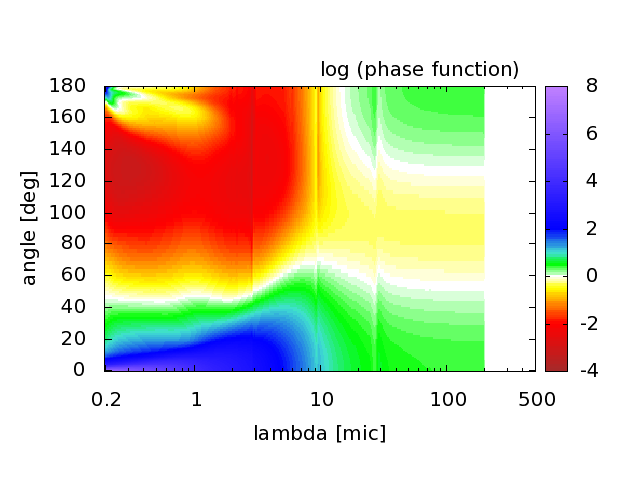}
}
\vspace{-5mm}
\centerline{
\includegraphics[angle=0,width=8.cm]{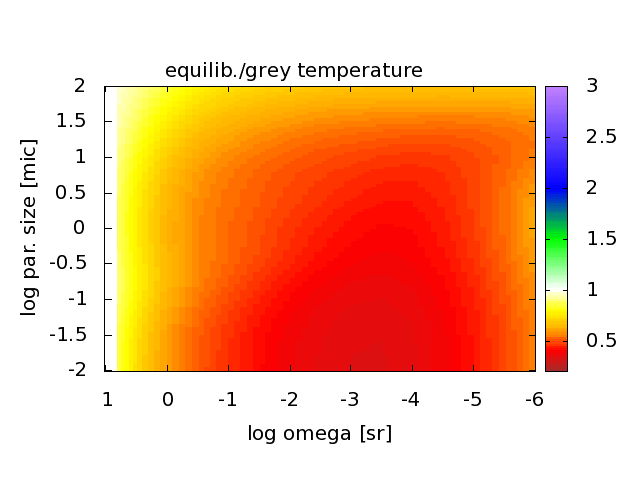} 
\includegraphics[angle=0,width=8.cm]{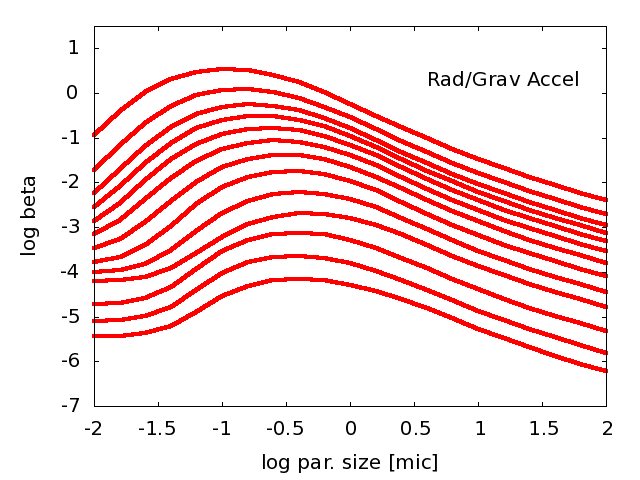}      
}
\caption{Ammonia.    
Pictures and notation are the same as in the previous figure.
}
\label{ammonia} 
\end{figure*}

\end{document}